\documentclass[review,3p,11pt,a4paper,times]{elsarticle}

\usepackage{times}

\newcounter{lastnote}

\usepackage{rotating}
\usepackage{ulem}
\usepackage{amssymb,amsmath,amsthm,amsfonts,bm,float}
\usepackage{textcomp}
\setcitestyle{authoryear,open={(},close={)}}

\usepackage{optidef}
\usepackage{subcaption}
\usepackage{xcolor}
\usepackage{multirow}
\usepackage[ruled,vlined]{algorithm2e}
\usepackage{url}
\usepackage{ulem}

\def\BibTeX{{\rm B\kern-.05em{\sc i\kern-.025em b}\kern-.08em
    T\kern-.1667em\lower.7ex\hbox{E}\kern-.125emX}}

 \definecolor{darkolivegreen}{rgb}{0.33, 0.42, 0.18}
 \definecolor{crimson}{rgb}{0.86, 0.08, 0.24}
\DeclareMathOperator*{\argmin}{arg\,min}

\begin{document}
\journal{XXXX}

\begin{frontmatter}


\title{Modeling Atmospheric Data and Identifying Dynamics \\
  \large Temporal Data-Driven Modeling of Air Pollutants }



\author[Javier]{Javier Rubio-Herrero
\corref{cor1}}
\ead{javier.rubioherrero@unt.edu}
\author[Carlos]{Carlos Ortiz Marrero}
\ead{carlos.ortizmarrero@pnnl.gov}
\cortext[cor1]{Corresponding author}
\author[Louis]{Wai-Tong Louis Fan\fnref{Louis2}}
\ead{lfan@cmsa.fas.harvard.edu}
\address[Javier]{Department of Information Technology and Decision Sciences, G. Brint Ryan College of Business, University of North Texas, 1155 Union Circle, Denton, TX 76201}
\address[Carlos]{Data Sciences and Analytics, Pacific Northwest National Laboratory, 902 Battelle Blvd, Richland, WA 99354}
\address[Louis]{Department of Mathematics, Indiana University, 831 East 3rd St. Bloomington, IN 47405}
\address[Louis2]{Center of Mathematical Sciences and Applications, Harvard University, Cambridge, MA 02138}

\date{}




\baselineskip24pt




\begin{abstract}
Atmospheric modeling has recently experienced a surge with the advent of deep learning. Most of these models, however, predict concentrations of pollutants following a data-driven approach in which the physical laws that govern their behaviors and relationships remain hidden. With the aid of real-world air quality data collected hourly in different stations throughout Madrid, we present an empirical approach using data-driven techniques with the following goals: (1) Find parsimonious systems of ordinary differential equations via sparse identification of nonlinear dynamics (SINDy) that model the concentration of pollutants and their changes over time; (2) assess the performance and limitations of our models using stability analysis; (3) reconstruct the time series of chemical pollutants not measured in certain stations using delay coordinate embedding results. Our results show that Akaike's Information Criterion can work well in conjunction with best subset regression as to find an equilibrium between sparsity and goodness of fit. We also find that, due to the complexity of the chemical system under study, identifying the dynamics of this system over longer periods of time require higher levels of data filtering and smoothing.
Stability analysis for the reconstructed ordinary differential equations (ODEs) reveals that more than half of the physically relevant critical points are saddle points, suggesting that the system is unstable even under the idealized assumption that all environmental conditions are constant over time.

\end{abstract}
  

\end{frontmatter}


\section{Introduction}\label{sec:intro}
Chemists and environmental scientists refer to nitrogen oxides as the group of compounds that contain nitrogen and oxygen. The most important of those gases, nitric oxide ($NO$) and nitrogen dioxide ($NO_2$), are of special interest because they are byproducts of many human activities. In 2013, $76.4\%$ of the tropospheric $NO_x$ had an anthropogenic source. Of that total, $75.5\%$ originated from fossil fuel combustion and industrial processes \citep{IPCC2013}.

The relevance of these two gases in how ozone ($O_3$) is distributed on earth makes the control of their emissions of paramount importance from a policy, environment, and health perspective. The chemical reactions that lead to the creation of ozone and the connection of this phenomenon with urban pollution have been largely studied \citep{crutzen1970influence,crutzen1979role,seinfeld2016atmospheric}.

Ozone is critical in the stratosphere, as it protects the earth from the harmful effects of UV radiation by absorbing it.  However, the creation of ozone in the troposphere poses a serious pollution problem as it is the main component of smog and thus becomes part of the air we breathe \citep{blaszczak1999nitrogen}. In addition, its creation is deeply affected by climate change, as the chemical reactions that produce it are very sensitive to temperature and lead to higher concentrations at higher temperatures \citep{aw2003evaluating}.

In order to fight against the effects of pollution, many governments have implemented policies at different levels, namely, national, regional, and local. 
For example, parts of California have seen a significant decrease in pollution over a period of 20 years. More recently, the Chinese government imposed strict measures to fight against the spread of COVID-19 \citep{zhu2020novel}, which resulted in a cease of transportation and industrial activities in most parts of the country. A collateral effect of these strict measures was a sharp decrease in the emissions of $NO_2$.

At a council level, some major capitals are introducing their own control policies. In an attempt at pedestrianization, many councils are moving towards transit models that hamper the use of cars in densely populated urban areas. In 2018, Spain's capital, Madrid, designated parts of its downtown as low-emission zones, known as \textit{Madrid Central} (MC) \citep{Madrid_Central}. This city's council limited the access of certain vehicles, albeit the transit of residents' vehicles remained permitted. Recent research shows that these limitations have decreased the concentration of $NO_2$ in downtown \citep{lebrusan2019assessing}. 

\section{Literature Review and Objectives}\label{sec:lit_review}

Predicting the concentration of pollutants in the atmosphere is a well-studied topic \citep{cooper1997survey, daly2007air}. In \cite{li2016deep} the authors rightfully indicate that there are two types of modeling efforts when it comes to forecasting the concentration of pollutants in the atmosphere. On the one hand, researchers can tackle this problem under a deterministic approach, in which the physics of the model comes into play in the form of diffusion equations, the pollutants' chemical characteristics, or fluid dynamics. These models usually require the solution of nonlinear mathematical relationships typically expressed in the form of partial differential equations \citep{ogura1962scale, wilhelmson1972pressure, lanser1999analysis}. Other physics-based models compute air parcels and track (or backtrack) the dispersion of atmospheric pollutants \citep{stein2015noaa}.

On the other hand, purely data-driven, statistical approaches bypass the physics that underlie the complicated behavior of these pollutants. The range of tools employed in these approaches vary considerably. For example, classic statistical models were employed by \cite{robeson1990evaluation}, who used univariate deterministic/stochastic models, ARIMA models, and bivariate models to forecast maximum ozone concentrations. With this same goal, \cite{prybutok2000comparison} proposed a regression model, a Box-Jenkins model, and a fully-connected neural network and concluded that the neural network performed better. Simulation models have also been an alternative in this context: an example is \textit{CALIOPE} \citep{baldasano2011annual}, a Spanish air quality forecasting system for temporal and spatial prediction of gas phase species and particulate matter (i.e., $NO_2, SO_2, O_3$ and $PM_{10}$). Finally, the advent of deep learning brought new opportunities for more accurate forecasting. Recurrent neural networks (RNNs) in general, and \textit{Long Short-Term Memory} (LSTMs) in particular, have been explored profusely as a means to explain the evolution of unknown variables over time. A recent example is the work by \cite{feng2019recurrent}. More recently, other black-box-based models were used to forecast  concentrations of pollutants or air quality indexes
\citep{abirami2021regional,zhang2020multi,liu2020study}. LSTMs were used by \cite{pardo2017air} to predict the levels of $NO$ and $NO_2$ with lags of 8, 16, and 24 hours, producing results that were superior to those reported by \cite{baldasano2011annual}.

 These different approaches within data-driven solutions present advantages and disadvantages. Neural networks require many training examples to attain accurate results and do not extrapolate well outside the regime in which they where trained \citep{bilbrey2020tracking}. Most importantly, these black box models produce uninterpretable relationships between the variables. Conversely, more traditional statistical procedures force the selection of a model or a family of models prior to calibrating and estimating coefficients, thus reducing the modeling flexibility. However, they provide clear, closed-form mathematical expressions that relate dependent and independent variables.


To address some of the challenges outlined above, some researchers have started to look at ways to integrate the data-driven efforts with domain modeling (i.e., modeling that aims at tuning parameters that set the relationships between variables and that also enforces the laws that govern the systems under consideration). Most of the work is beginning to coalesce under the banner of \textit{Scientific Machine Learning} \citep{baker2019workshop} and promising approaches are continually being developed and improved \citep{raissi2019, brunton2016discovering, chen2018neural, rackauckas2020universal}. 
Inspired by these recent successes in the field, the goal of this article is to outline an empirical data-driven, but domain-aware, framework to model atmospheric pollutants. Such a framework bridges the gap between these two classic perspectives (deterministic and data-driven) and provides methods that can leverage data as well as produce relationships between atmospheric chemical species that capture the physics of the system being modeled. In this paper, we apply these techniques to find a system of ordinary differential equations (ODEs) from real-world measurements of $NO_2$ and $O_3$. 

As previously discussed, the use of these techniques have clear applications in policy-making environments at the national, regional, and local levels where accurate quantitative tools are of vital importance to assess atmospheric contamination. In turn, these tools can help policy makers determine the benefits and impact of their pollution control techniques \citep{popp2006international}. The structure of this paper is as follows:
\begin{enumerate}
    \item In Section \ref{sec:SINDy} we propose an alternative optimization approach for sparse identification of nonlinear dynamics (SINDy) \citep{brunton2016discovering}. We will apply this approach to real time-series data collected in various air quality stations located in Madrid in order to find systems of ordinary differential equations (ODEs) that will capture the dynamics of ozone and nitrogen dioxide at those geographical spots for a given time frame. We will discuss the implications of noisy datasets in this context and how that impacts the performance of SINDy.
    \item 
    In Section \ref{sec:properties_ODEs}, we analyze some basic mathematical properties of the ODEs reconstructed from the data in Section \ref{sec:SINDy} and offer some insight regarding the global behavior of the dynamics of the concentrations of $NO_2$ and $O_3$. For each air quality station that measures both chemical species, we classify the critical points of the system of ODEs according to a stability analysis. The goal of this analysis is to provide us with a way to interpret the performance and limitations of our fitted equations.
    \item In Section \ref{sec:reconstruction} we reconstruct the time series of the concentrations of $O_3$ at those stations where only $NO_2$ readings are available. To this end, we rely on \textit{Takens' embedding theorem} to perform the reconstruction. The goal of this reconstruction is to provide the foundation of a method that allows to identify the dynamics of a chemical species in a location where readings of this species are not available.
\end{enumerate}

\section{SINDy for Atmospheric Data}\label{sec:SINDy}
\subsection{Motivation}\label{subsec:SINDy_motivation}

The application in hand plays a crucial role in the dichotomy between flexibility and tractability posed by black-box models and regression models. Many dynamical systems can be explained with low-order mathematical expressions. As a matter of fact, the dynamics of many physical and chemical systems are modeled as systems of ODEs. In particular, in the field of atmospheric chemistry many chemical reactions and interactions that take place on the Earth's atmosphere can be modeled this way under certain conditions.

Chemical reactions are governed by their rate equations. The kinetics of a chemical reaction show how the concentrations of the reactants and the products vary during the reaction. Rate equations are easy to obtain from elementary reactions in closed systems (i.e., in systems where only those reactions occur and there is not a flux of other molecules entering or exiting those systems). For instance, \textit{mass-action kinetics} suggest that the reaction
\begin{align*}
    aA+bB\rightarrow cC+dD
\end{align*}
\noindent consumes reactant $A$ at a rate 
of $mole/(m^3\cdot s)$ 
given by the differential equation
\begin{align*}
    \frac{dA}{dt}=-k[A]^a[B]^b,
\end{align*}
\noindent where $k$ is the rate constant of the reaction \citep{erdi1989mathematical}. 
In turn, the exponents of the concentrations, $a$ and $b$, called the \textit{partial orders} of the reaction are in this case the stoichiometric coefficients of the chemical reaction. Their sum is referred to as the \textit{overall order} of the reaction. 

Eventually, in a closed system, the reactants are exhausted. In open systems with influx and efflux of several chemical species, like the troposphere, many chemical reactions occur simultaneously. Consequently, molecules are constantly created and destroyed, either created or consumed as a result of those reactions. This phenomenon yields a steady-state in the concentrations that stems from a dynamic equilibrium \citep{denbigh1948kinetics}.  Also, in open systems the reactions occur in multiple steps and the partial orders of rate equations usually do not match the stoichiometric coefficients. Moreover, since a reactant can be part of several reactions at the same time (i.e., can be consumed or produced as a result of other reactions), the time-series of its concentration is not necessarily given by the the rate equation of a single reaction.

As an example, consider the \textit{Leighton cycle} \citep{leighton2012photochemistry} that explains the formation of ozone from nitrogen oxides in the troposhere in unpolluted conditions:
\begin{eqnarray}
NO_2 + hv&\xrightarrow{J_t}& NO+O(^3P),\label{eqn:leighton_1}\\
O(^3P) + O_2 &\rightarrow& O_3,\label{eqn:leighton_2}\\
NO + O_3&\xrightarrow{k_3}& NO_2 + O_2,\label{eqn:leighton_3}
\end{eqnarray}
\noindent where $hv$ represents energy from solar radiation (as calculated by the product of Planck's constant, $h$, and the frequency of the wave of solar radiation, $v$) and $O(^3P)$ denotes an oxygen atom in its fundamental state. In turn, $k_3$ is the rate constant of the third reaction and the reaction rate $J_t$ represents the actinic flux, which varies over time and depends largely on the incidence of photons, thus presenting different values according to other factors such as cloud cover or season. The reactions (\ref{eqn:leighton_1})-(\ref{eqn:leighton_3}) present the following kinetics \citep{marsili1996simplified}:
\begin{eqnarray}
\frac{d[NO_2]}{dt}&=&-J_t[NO_2]+k_3[NO][O_3],\label{eqn:kinetics_NO2}\\
\frac{d[O_3]}{dt}=\frac{d[NO]}{dt}&=&J_t[NO_2]-[NO][O_3].\label{eqn:kinetics_O3}
\end{eqnarray}

In unpolluted conditions, the ozone present in the troposphere is due to transport from the stratosphere and photochemical production. Its destruction is also due to photochemical reactions and from deposition on the earth's surface. These processes happen at a rate that maintains the level of ozone approximately constant in this layer of the atmosphere and, in these circumstances, the above represents a null cycle in which there is not any net production nor destruction of these chemical species. Therefore, their kinetics reach a pseudo-steady-state that can be expressed as
\begin{eqnarray*}
\frac{J_t}{k_3}&=&\frac{[NO][O_3]}{[NO_2]}.
\end{eqnarray*}

This relationship explains why during daylight hours, when the actinic influx is large, there is an increment of the concentrations of $NO$ and $O_3$ at the expense of a destruction of $NO_2$ \citep{national1992rethinking}. It also explains why this trend is reversed during the night hours. 

The kinetics in \textit{polluted} conditions turn out to be much more complex. The only known source for the creation of $O_3$ is via the photolysis of $NO_2$ (see Equation (\ref{eqn:leighton_1})). Thus, the ozone build-up that appears in those conditions is due to an excess of $NO_2$ produced by man-made pollution. Indeed, pollution is responsible for the presence of free radicals that initiate a series of chain reactions that lead to the creation of more $NO_2$. The basic Leighton cycle does not capture these side reactions and, consequently, it is disrupted and results in a net creation of $O_3$. For this reason, characterizing a polluted environment requires the addition of the effect of those free radicals, often difficult to measure, which culminates in a much more complex and intertwined series of reactions. Readers interested in further details on the chemical reactions that take place in such environments may refer to \cite{finlayson1986atmospheric}.

In spite of the inherent complexity in modeling the dynamics of ozone and nitrogen oxides in urban environments, Marsili-Libelli (1996) simplified considerably their kinetics by including the concentrations of the free radicals that disrupt the Leighton cycle into the kinetic rates of basic chemical reactions. His results were satisfactorily tested with data from a real urban environment in Italy. The resulting model expressed the variation of $NO_2$ and $O_3$ in terms of polynomials of order $2$ of the concentrations of the chemical species involved and the kinetic rates were calculated by calibration with an adaptive polyhedron search. Following the insight provided by Marsili-Libelli (1996) that this complex system can be simplified with a system of low-order ODEs that include the elusive information from free radicals into the kinetic rates, we explore the possibility of modeling the dynamics of chemical species in polluted environments similarly to their kinetics in unpolluted cases (equations (4) and (5)). We do this with a data-driven method that incorporates implicitly the complexity introduced by the side reactions of the free radicals into the kinetic rates. Hence, we find our research question in how we can 
develop data-driven methods that are able to capture the dynamics of a complex, atmospheric open system in a closed mathematical form by identifying the values of the coefficients of the rate equations corresponding to some of the reactions that occur. 
In our empirical work, we focus on the dynamics of two pollutants, namely nitrogen dioxide and ozone, whose time-varying concentrations are interdependent.

As mentioned in Section \ref{sec:lit_review}, there have been attempts to predict the concentration of ozone in the atmosphere. Our approach differs from all these in that we propose a regression approach that can capture the dynamics of the atmosphere in a way that different chemical species and their concentrations over time are interrelated, thus offering a closed form of the rate equations that govern the chemical reactions occurring during the selected time frame. In addition, in Sections \ref{sec:properties_ODEs} and  \ref{sec:reconstruction} we will use those governing equations to offer analytical insight of the dynamics of these species and to reconstruct the time series of ozone in those stations that do not measure them.

\subsection{Mathematical representation of the system dynamics}\label{subsec:math_representation}
Our goal is to find a series of ODEs that describe the chemical dynamics in the troposphere. That is, a system of the form,
\begin{align}
    \mathbf{\dot{y}}(t)=\mathbf{F}(\mathbf{y}(t)),\quad i=1,2,\dots,p,
    \label{eqn:ODEs}
\end{align}
\noindent where $\mathbf{y}(t)=(y_1(t),y_2(t),\dots,y_p(t))^T\in\mathbb{R}^p$ is a vector containing the time response of the concentrations of the $p$ chemical species under study, $\mathbf{\dot{y}}(t)=(\dot{y}_1(t),\dot{y}_2(t),\dots,\dot{y}_p(t))^T$ is the vector of derivatives, and $\mathbf{F}(\mathbf{y}(t))$ is a vector field acting on $\mathbf{y}(t)$. In \cite{brunton2016discovering} the authors outlined a methodology to estimate the functional form $\mathbf{F}$ given samples from $\mathbf{y}$ and $\mathbf{\dot{y}}$ by solving the a series of least squares problems, 
\begin{align}
    \min_{\boldsymbol{\beta_i}} ||\mathbf{\dot{\tilde{y}}_i}-\mathbf{\tilde{F}}\boldsymbol{\beta_i}||_2^2,\quad i=1,2,\dots,p.
    \label{eqn:regression}
\end{align}
where $\mathbf{\dot{\tilde{y}}_i}=(\dot{y}_i(t_1), \dot{y}_i(t_2),\dots, \dot{y}_i(t_m))^T\in\mathbb{R}^m$ is a vector of observed derivatives and $\boldsymbol{\beta_i}\in\mathbb{R}^{n+1}$ is the vector of regression coefficients. The matrix $\mathbf{\tilde{F}}\in\mathbb{R}^{m\times (n+1)}$ contains information about candidate nonlinear basis functions (plus the intercept term) over a time horizon $m\gg n$. For example, if we have a system of two chemical species ($p=2$) and we assume that the entries of $\mathbf{F}(\mathbf{y}(t))$ are composed of second-order polynomials over $y_1(t)$ and $y_2(t)$ (i.e., $n=5$), then
\begin{align*}
\mathbf{\tilde{F}}= \begin{bmatrix}
1 & y_1(t_1) & y_2(t_1) & y^2_1(t_1) & y^2_2(t_1) & y_1(t_1)y_2(t_1)\\
1 & y_1(t_2) & y_2(t_2) & y^2_1(t_2) & y^2_2(t_2) & y_1(t_2)y_2(t_2)\\
\vdots & \vdots & \vdots & \vdots & \vdots & \vdots\\
1 & y_1(t_m) & y_2(t_m) & y^2_1(t_m) & y^2_2(t_m) & y_1(t_m)y_2(t_m)\\
\end{bmatrix}.    
\end{align*} 

Notice that the solution to the minimization problem (\ref{eqn:regression}) yields an approximate solution to $\mathbf{F}$ in \eqref{eqn:ODEs}. Also note that when data from $\mathbf{\dot{\tilde{y}}_i}$ are not available, we can approximate these vectors via numerical differentiation using the data vector $\mathbf{\tilde{y}_i}=(y_i(t_1), y_i(t_2),\dots, y_i(t_m))$, as suggested by \cite{kaiser2018sparse}. In particular, we calculated each derivative as $\dot{\tilde{y}}_i(t_j)=\frac{\tilde{y}_i(t_j)-\tilde{y}_i(t_{j-1})}{t_j-t_{j-1}}$. With our data being collected on an hourly basis (see Subsection \ref{subsec:data}), this numerical differentiation reduces to calculating $\dot{\tilde{y}}_i(t_j)=\tilde{y}_i(t_j)-\tilde{y}_i(t_{j-1})$. As we will also mention in that subsection, these computations take place after a filtering and splining process aimed at reducing data noise. Other approaches to curb the effect of noisy data on numerical differentiation could also be possible \citep{chartrand2011numerical} and would be an interesting area to explore in the future.

In most occasions not all the terms in the chosen linear functional form are needed, as it is very likely that some are not relevant for explaining the dependent variable in question. Therefore, a frequent subproblem within regression is that of finding the subset of terms that provides the best representation of the independent variable. This search for a more parsimonious or sparse mathematical expression is the core idea explored in \cite{brunton2016discovering}. We analyzed the advantages and drawbacks of \textit{LASSO} regression \citep{tibshirani1996regression} and \textit{best subset} regression \citep[Chapter~19]{hill2006statistics} and concluded that for our application the latter was a better choice to obtain accurate representations of the dynamics of $NO_2$ and $O_3$ at different geographical points of Madrid. A description of both methods is detailed in \ref{sec:methods_considered}.  

\subsection{Method selected}\label{subsec:Method_selected}
The literature on systems identification is vast and its applications have been studied for many years \citep{ljung1994modeling,ljung1999system}. However, the application of LASSO regression to \textit{sparse identification of nonlinear dynamics} (SINDy) is more recent and, although some researchers proposed its use in this context shortly after LASSO was developed \citep{kukreja2006least}, it became more prominent in the literature since \cite{brunton2016discovering}. Consequently, there have been multiple efforts to find sparse representations of physical and biological systems as well as population dynamics \citep{mangan2016inferring,kaiser2018sparse}. As far as chemical systems are concerned, \cite{bhadriraju2019machine} modeled the dynamics of a continuous stirred tank reactor with an adaptive sparse identification method that involved sparse model identification, re-estimation of regression coefficients, and stepwise regression. These same authors also tackled the same problem with SINDy in conjunction with a neural networks controller that determined when the outputted equations needed to be re-evaluated \citep{bhadriraju2020operable}. \cite{hoffmann2019reactive} extended SINDy with ansatz functions to describe what they called ``reactive SINDy" in order to eliminate the spurious reactions that are typically captured in reaction networks behind biological processes. The applicability of SINDy and \textit{symbolic regression} (SymReg) for predicting the dynamics in a distillation column within the context of a manufacturing process was put to the test by \cite{subramanian2021white}. The authors found that SINDy performed better than SymReg and was able to identify terms related to \textit{Fick's law} and \textit{Henry's law}. They concluded that all the dynamics present in that system could not be captured by only one method and suggested the parallel use of different machine learning algorithms to capture the system's complexity entirely. A very recent and promising attempt to reduce the complexity inherent to solving the systems of ODEs that stem from large chemical networks was presented by \cite{grassi2021reducing}. In this case, however, the authors did not resort to a method like SINDy, but rather proposed a combination of encoders and decoders that produced a transparent and interpretable latent state with many less variables. The problem then was the correct definition of the topology of the compressed chemical network that eventually produced a result that had to be mapped back to the original set of variables. In \cite{narasingam2018data}, the authors successfully demonstrated how to construct reduced order models using SINDy to approximate the dynamics of a nonlinear hydraulic fracturing process.

The tractability considerations discussed in \ref{sec:methods_considered} suggested that LASSO regression was a sensible option for identifying the dynamics of chemical species in the troposphere. However, the aforementioned works found sparse identifications of dynamic systems from data that were previously generated. During the course of our present research it was our experience that LASSO does not perform well in the context of SINDy when dealing with real-world data from multiple geographical locations, even after the stage of data preprocessing. We attribute this to the considerable amount of noise present in the data and the open nature of the system we are trying to model. While LASSO certainly could find sparse systems of ODEs, trying to solve numerically the systems of ODEs denoted by (\ref{eqn:regression}) was very problematic because of numerical issues caused by singularities.

Assuming that, as mentioned in Subsection \ref{subsec:SINDy_motivation}, the dynamics of the atmosphere can be explained with low-order polynomials, and based on equations (\ref{eqn:kinetics_NO2}) and (\ref{eqn:kinetics_O3}) for unpolluted environments, we conjectured that second-order polynomials would suffice to identify the dynamics of this system under polluted conditions. When dealing with two chemical species, $NO_2$ and $O_3$ (i.e., $p=2$), such polynomials have at most $5$ variables and there are $2^5=32$ possible regressions for each equation in (\ref{eqn:regression}). In these circumstances, brute-force enumeration of all the possible regressions is computationally affordable and therefore we opted for a best subset regression approach in which for each chemical species $i$, the vector of optimal regression coefficients $\boldsymbol{\hat{\beta}_i}$ was selected according to the \textit{Akaike information criterion} (AIC) \citep{akaike1998information}. We thus solved the problem
\begin{mini}|l|[1]
  {\boldsymbol{\beta_i}}{m\log{\left(\frac{\|\mathbf{\dot{\tilde{y}}_i}-\mathbf{\tilde{F}}\boldsymbol{\beta_i}\|_2^2}{m}\right)}+2\|\boldsymbol{\beta_i}\|_0}{}{BS(i):}
  {\label{eqn:best_subset_AIC}}
\addConstraint{\|\boldsymbol{\beta_i}\|_0}{\le n,}{}
\end{mini}
where the $\ell_0$-norm denotes the number of non-zero elements of the vector $\boldsymbol{\beta_i}$ (excluding the intercept). Selecting the AIC was motivated by the need of an equilibrium in the bias-variance tradeoff. This criterion seeks that equilibrium by taking into account both the sum of the square of the errors (hence tackling underfitting) and the number of variables in the regression (hence tackling overfitting). This was useful to compare all the possible models and allowed us to implement an algorithm that could bypass the numerical issues encountered when using the ODE solvers (see \ref{sec:appendix_algorithm}). Our algorithm was programmed with \textit{MATLAB R2020a} and was built on the structure of the code developed in \cite{brunton2016discovering}. This code was adapted to solve the systems of ODEs with the output of the regressions obtained from the best subset method. MATLAB has different ODE solvers that can be used in various environments. In our particular case, we found that many of the systems of ODEs produced were stiff, and thus we used MATLAB's stiff solver \textit{ode15s}. This improved the integrability of the systems of ODEs, although still presented numerical problems in some occasions. These problems stemmed from singularities or regions were the derivative changed very rapidly. To address this problem, we introduced in our algorithm an upper bound for the values of the derivative in such a way that those regressions that eventually led to those ill-posed solutions would be discarded.
 
 In the absence of numerical problems, our algorithm returns the optimal solution to Problem \eqref{eqn:best_subset_AIC} for each $i=1,2,\dots,p$ and provides representations of (\ref{eqn:regression}) that minimize the AIC. In the presence of numerical problems, it finds representations of (\ref{eqn:regression}) that minimize the AIC while also being numerically tractable.


\subsection{Data}\label{subsec:data}

\begin{figure}[h!]
  \centering
    \includegraphics[scale=0.2]{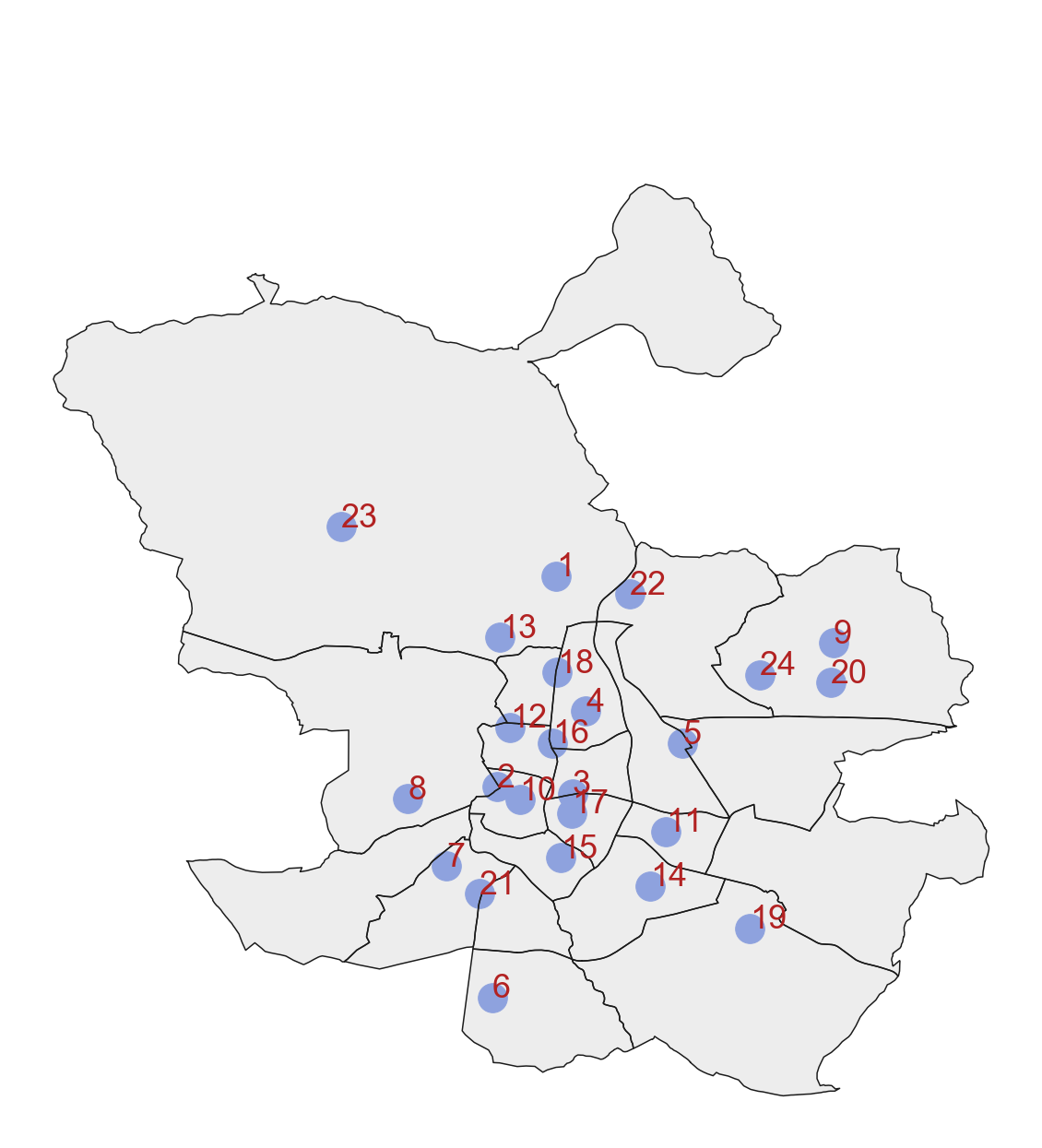}
    \caption{Stations for pollution control in Madrid}
    \label{fig:Stations_Plot}
\end{figure}

The data used in this study were collected from \textit{Madrid's City Council Open Data website} \citep{Madrid_datos}. The authors selected this data set because of the vast and detailed amount of information that it provides, as it contains hourly readings between 2001 and 2018 of various pollutants in 24 different stations located across areas of downtown Madrid, as well as its outskirts (see Figure \ref{fig:Stations_Plot} and Table \ref{table:stations}). Amongst these pollutants, there are readings for $NO$, $NO_2$, $NO_x$, and $O_3$ in $\mu g/m^3$. All stations have full readings of $NO$, $NO_2$, and $NO_x$ for all hours. Only $14$ stations (marked in bold in Table \ref{table:stations}) have readings of $O_3$ for all hours; the remaining $10$ stations did not capture readings of $O_3$. For this reason, our regressions were conducted for only the 14 stations for which both $NO_2$ and $O_3$ data were available. In Section \ref{sec:reconstruction}, we will discuss how we used our results in these stations to reconstruct the time series of $O_3$ readings in the other 10 stations. 

\begin{table}[h!]
\centering

\begin{tabular}{ |c|c||c|c| } 

 \hline
 \textbf{Number} & \textbf{Station name} & \textbf{Number} & \textbf{Station name} \\ 
 \hline\
1 & Pza. de Espa\~na & 13 & Vallecas\\ 
2 & \textbf{Escuelas Aguirre} & 14 & Mendez Alvaro\\ 
3 & Avda. Ram\'on y Cajal & 15 & Castellana\\ 
4 & \textbf{Arturo Soria} & 16 & \textbf{Parque del Retiro}\\
5 & \textbf{Villaverde} & 17 & Plaza Castilla\\
6 & \textbf{Farolillo} & 18 & \textbf{Ensanche de Vallecas}\\
7 & \textbf{Casa de Campo} & 19 & Urb. Embajada\\
8 & \textbf{Barajas Pueblo} & 20 & \textbf{Pza. Fern\'andez Ladreda}\\
9 & \textbf{Pza. del Carmen} & 21 & Sanchinarro\\
10 & Moratalaz & 22 & \textbf{El Pardo}\\
11 & Cuatro Caminos & 23 & \textbf{Juan Carlos I}\\
12 & \textbf{Barrio del Pilar} & 24 & \textbf{Tres Olivos}\\
 \hline
\end{tabular}
\caption{List of stations for pollution control in Madrid (in \textbf{bold} those that measured $O_3$).}
\label{table:stations}
\end{table}

As mentioned in the previous section, the collected raw data came from sensors, which are naturally noisy \citep{Madrid_datos}. In the presence of such noise, it is very complicated to find systems of ODEs that that can be solved numerically without issues. Therefore some data preprocessing was needed. The following summarizes the operations performed on the set of raw data:
\begin{itemize}
    \item \textbf{Data normalization:} The order of magnitude of the concentrations of the different chemical species in the atmosphere may differ greatly. For this reason, the time series of all molecules were standardized (i.e., for each data point of the time series we subtracted the average concentration during the time frame considered and divided over the standard deviation). It was our experience that this lead to fewer numerical errors when we integrated the systems of ODEs represented by equation (\ref{eqn:regression}). We will denote our original $M$ data samples of normalized concentrations of $p$ chemical species as $\mathbf{\tilde{w}_i}, i=1,2,\dots,p$. 
    
    \item \textbf{Data filtering:} the excess of data noise made it difficult to extract trends in the concentrations of pollutants over time. In order to address this issue we perform a Gaussian-weighted moving average filter over our data. This filter, as implemented in MATLAB, uses a window size determined heuristically that is attenuated according to a smoothing parameter $\alpha\in[0,1]$. Values of $\alpha$ close to $0$ reduce the window size (i.e., reduces the smoothing), whereas values of $\alpha$ close to $1$ increase the window size (i.e., increases the smoothing). In very noisy and long time series, high values of $\alpha$ might be needed to extract meaningful patterns. However, this might come at the cost of excessive damping and the resulting time series might not be a good representation of the underlying data set. This parameter will be critical in our experiments, as we will discuss in Subsection \ref{subsec:results}.
    
    \item \textbf{Data splining:} raw data were collected hourly, but the integration of a continuous-time system of ODEs requires a finer discretization of the time space. In consequence, and after attempting finer and coarser discretizations, we proceeded with the creation of 100 points between each original pair of data points, following a \textit{modified Akima interpolation} (MAI) \citep{akima1970new}.  MAI helps with reducing excessive undulations that may occur with regular cubic splines. Figure \ref{fig:data_preprocessing} shows an example of two normalized time series that were filtered with different values of $\alpha$. Higher values of this parameter provide smoother but excessively dampened series. Lower values of $\alpha$ result in more realistic but also more unpredictable time series. In Figure \ref{fig:data_prep_a} the modified Akima interpolation does a very good job in avoiding undulations in the last four data points of the $NO_2$ series.
    
\begin{figure}[h!]
     \centering
     \begin{subfigure}[h!]{0.49\textwidth}
         \centering
         \includegraphics[width=\textwidth]{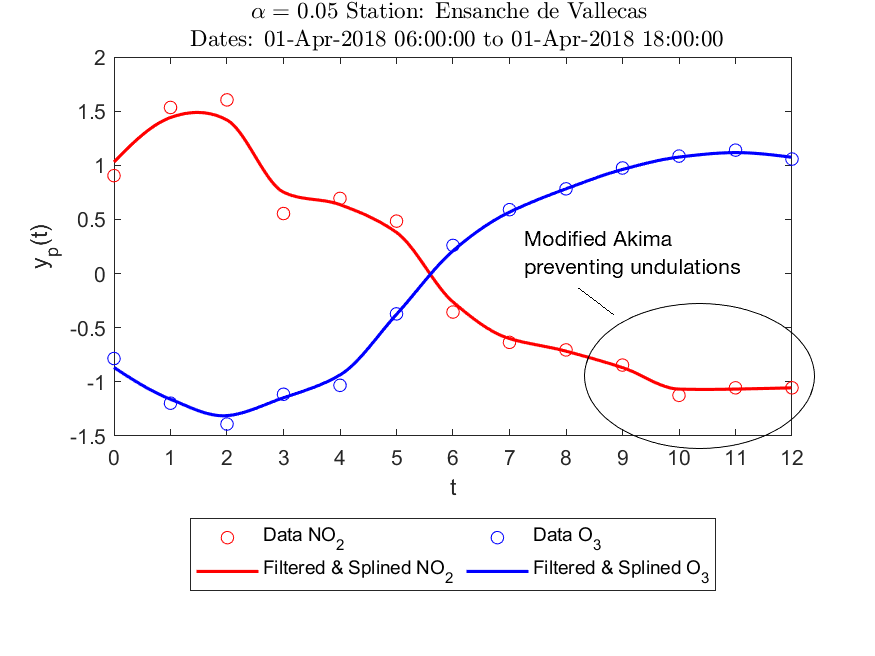}
         \caption{$\alpha=0.05$}
         \label{fig:data_prep_a}
     \end{subfigure}
     \hfill
     \begin{subfigure}[h!]{0.49\textwidth}
         \centering
         \includegraphics[width=\textwidth]{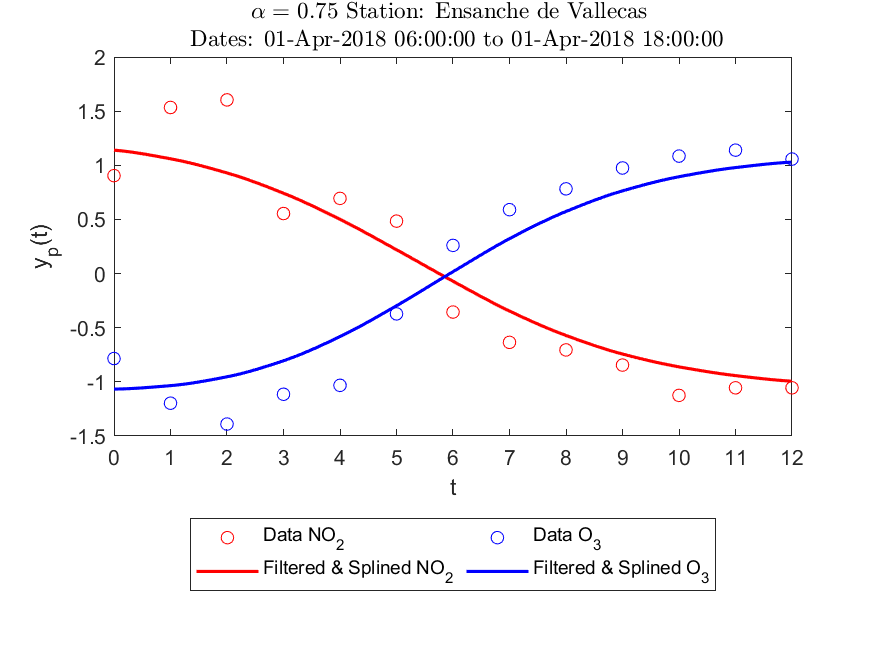}
         \caption{$\alpha=0.75$}
         \label{fig:data_prep_b}
     \end{subfigure}
        \caption{Effect of filtering and splining on normalized data}
        \label{fig:data_preprocessing}
\end{figure}
\end{itemize}

After filtering and smoothing, the resulting vector of normalized concentrations will be our $\mathbf{\tilde{y}_i}\in \mathbb{R}^m,i=1,2,\dots,p$. Note that, by means of the splining operation, these vectors have notably more time samples than the original normalized vectors $\mathbf{\tilde{w}_i},i=1,2,\dots,p$.  

\subsection{Optimal smoothing factor for regression}\label{subsec:results}
The goal of our best subset regression approach is not only to find the best system of differential equations in the sense of the AIC, but also that the solutions of those systems represent a good fit with respect to the raw data. The regression coefficients in (\ref{eqn:regression}) are found by solving Problem \eqref{eqn:best_subset_AIC} with the vectors $\mathbf{\tilde{y}_i},i=1,2,\dots,p$. This means that the suitability or goodness of the fitted regressions is measured against data that have been previously manipulated. An excellent fit of an overly manipulated time series will probably not be very useful in practical terms. However, a good fit of noisy raw data seems difficult to obtain, especially if we conjecture that the dynamics of this system can be modeled with a second-order polynomial.

In this context, the severity of the data filtering phase is paramount. It is sensible to develop a framework in which the hyperparameter $\alpha$ is tuned adequately. For a given time window $[t_0,t_f]$ that contains $M$ readings of $NO_2$ and $O_3$ in an air quality station $s$, let us define the \textit{root mean square error}
\begin{align}
    \text{RMSE}^{\alpha}_{i,s}=\sqrt{\frac{1}{M}\left(\mathbf{(\tilde{w}_{i}})_s-\mathbf{\hat{y}_{i,s}}(\alpha)\right)^2}.\label{eqn:RMSE}
\end{align}

In equation (\ref{eqn:RMSE}) the vector $(\mathbf{\tilde{w}_{i}})_s\in\mathbb{R}^M$ contains the original normalized $M$ readings of chemical species $i$ in station $s$. The vector $\mathbf{\hat{y}_{i,s}}(\alpha)\in\mathbb{R}^M$ contains the evaluations at those $M$ points performed after numerically solving the system of ODEs (\ref{eqn:regression}) when solving Problem \eqref{eqn:best_subset_AIC}. This way, each vector $\mathbf{\hat{y}_{i,s}}(\alpha)$ is compared to the original normalized observations. In order to find the smoothing parameter that performs best, a suitable approach is to find the solution to the following optimization problem:
\begin{equation}
\begin{array}{rrclcl}
\displaystyle \min_{\alpha}\max_{1\leq i \leq p} & \multicolumn{3}{l}{\displaystyle \overline{\text{RMSE}}^{\alpha}_i,}
\label{eqn:mix_max}
\end{array}
\end{equation}
\noindent where $\overline{\text{RMSE}}^{\alpha}_i$ is the average of $\text{RMSE}^{\alpha}_{i,s}$ over all the air quality stations. Therefore, Problem \eqref{eqn:mix_max} aims to calibrate the smoothing parameter $\alpha$ such that it minimizes the maximum forecasting error incurred, on average, by any chemical species. It is important to note that we should anticipate that the optimal value of $\alpha$ will be sensitive to the data selected for the analysis (i.e., it will be sensitive to location, number of time periods, etc.)

\begin{figure}[h!]
    \centering
    \includegraphics[scale=1.4]{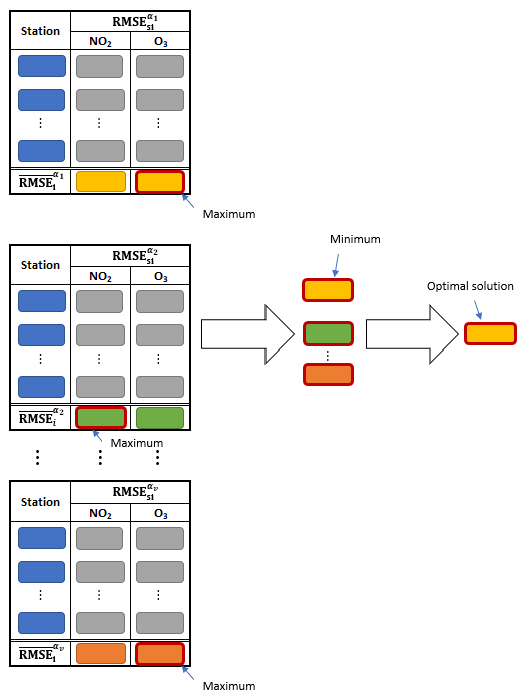}
\caption{Process for finding the solution to Problem \eqref{eqn:mix_max}}
\label{fig:min_max_process}
\end{figure}

The solution to this problem was tackled experimentally by considering $v$ different values for $\alpha$ such that $0<\alpha_1<\alpha_2<\dots<\alpha_v<1$. Figure \ref{fig:min_max_process} illustrates our procedure. For a given choice of $\alpha$ and for each station $s$, we solved problems $BS(i),i=1,2,\dots,p$ and found a system of $p$ (in our case $p=2$) differential equations for each air quality station that read both $NO_2$ and $O_3$. Once this system of ODEs was integrated numerically, we extracted the resulting vector $\mathbf{\hat{y}_{i,s}}(\alpha)$ of $M$ points and a value of $\text{RMSE}^{\alpha}_{i,s}$ was obtained by (\ref{eqn:RMSE}). Then, we repeated these operations for all stations and averaged those \text{RMSE}s to obtain $p$ different values of $\overline{\text{RMSE}}^{\alpha}_i$, whence the maximum was retrieved. After iterating over the $v$ different values of $\alpha$ considered, we selected the minimum of those maximum average \text{RMSE}s as the solution to Problem \eqref{eqn:mix_max}. In our experiments we filtered our time series with $v=21$ different values of $\alpha$, from $0.05$ to $0.95$ in intervals of $0.05$, plus $0.01$ and $0.99$. We did this for the $14$ stations that could read both $NO_2$ and $O_3$, which were selected as our chemical species ($p=2$).


\begin{figure}[h!]
  \centering
    \includegraphics[scale=0.34]{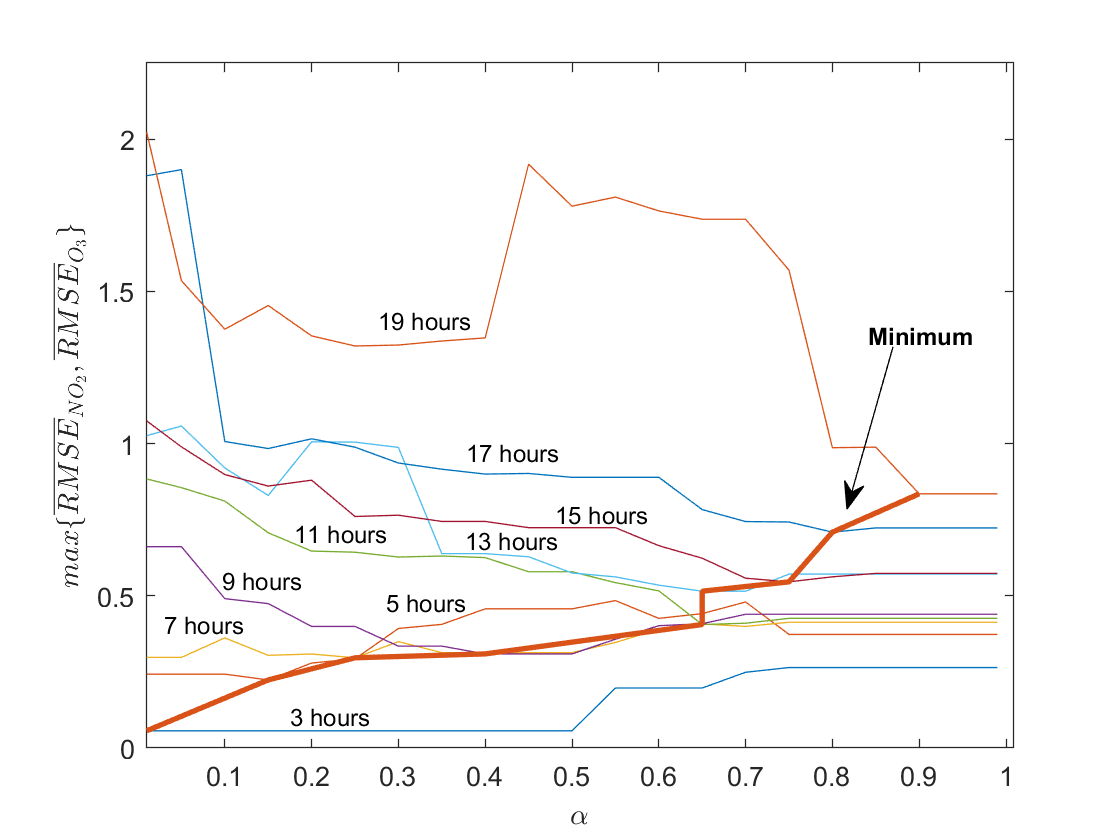}
    \caption{Worst average fit vs. smoothing factor}
    \label{fig:Fit_Smoothing_factor}
\end{figure}

Our results are consistent with the notion that larger time windows require a higher level of smoothing that can dampen the effect of noised data points. Consequently, it seems clear that the optimal value of $\alpha$ in Problem \eqref{eqn:mix_max} is nondecreasing as we enlarge our time window. The optimal value of the objective function in Problem \eqref{eqn:mix_max} is nondecreasing as well (i.e., the minimum of the maximum average RMSE does not decrease as we use larger data samples). This is shown in Figure \ref{fig:Fit_Smoothing_factor}, where we compared our results for a time window centered at noon on April 1\textsuperscript{st}, 2018, with increasing window sizes between 3 hours and 19 hours. 

\begin{figure}
     \centering
     \begin{subfigure}[h!]{0.49\textwidth}
         \centering
         \includegraphics[width=\textwidth]{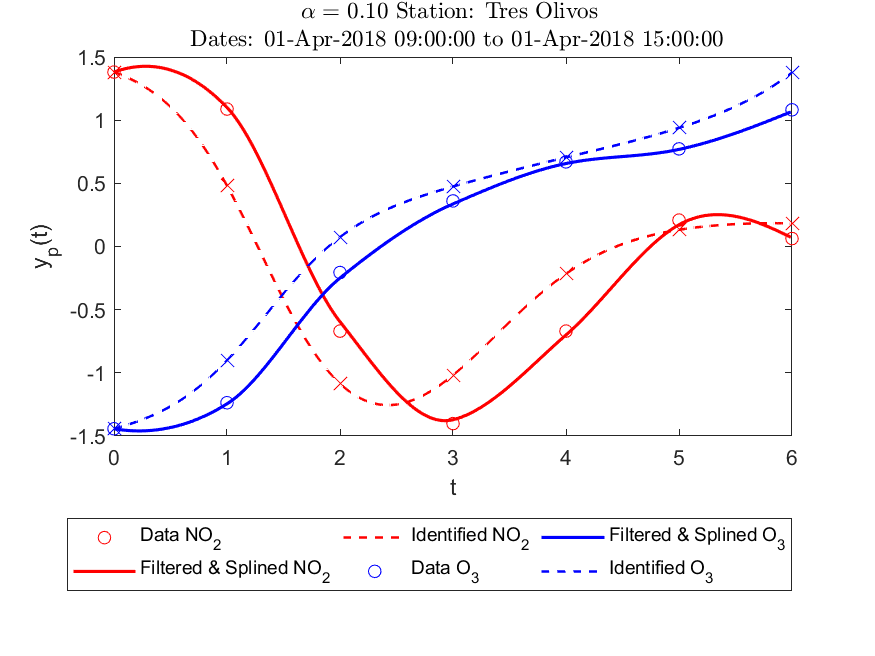}
         \caption{Time series, $\alpha=0.10$}
         \label{fig::tres_olivos_time_series_01}
     \end{subfigure}
     \hfill
     \begin{subfigure}[h!]{0.49\textwidth}
         \centering
         \includegraphics[width=\textwidth]{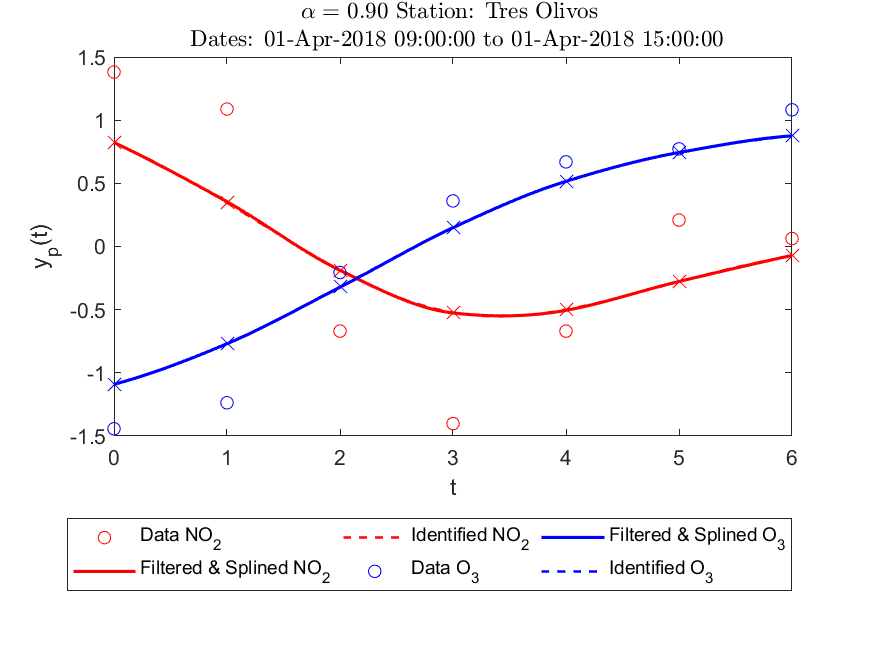}
         \caption{Time series, $\alpha=0.90$}
         \label{fig::tres_olivos_time_series_09}
     \end{subfigure}
     \hfill
     \begin{subfigure}[h!]{0.49\textwidth}
         \centering
         \includegraphics[width=\textwidth]{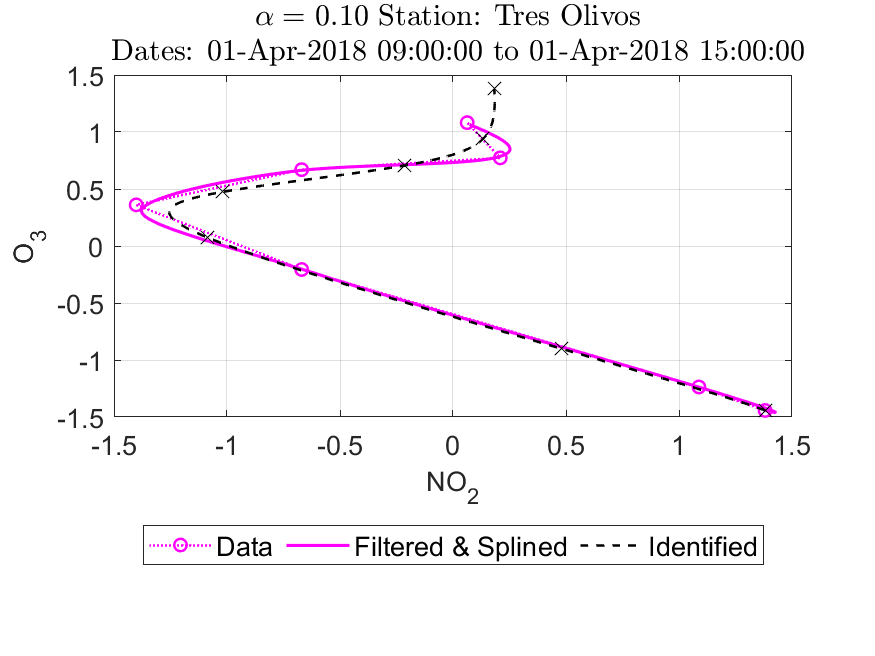}
         \caption{State diagram, $\alpha=0.10$}
         \label{fig::tres_olivos_state_01}
     \end{subfigure}
     \hfill
     \begin{subfigure}[h!]{0.49\textwidth}
         \centering
         \includegraphics[width=\textwidth]{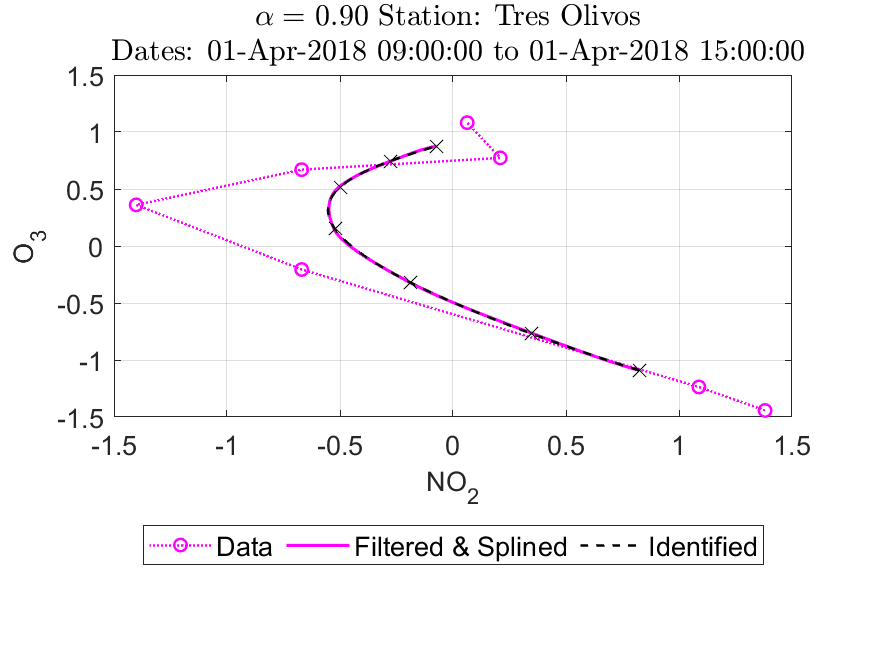}
         \caption{State diagram, $\alpha=0.90$}
         \label{fig::tres_olivos_state_09}
        \end{subfigure}
        \caption{Comparison of results with two disparate values of the smoothing factor.}
        \label{fig:tres_olivos}
\end{figure}

As already discussed, the usefulness of the regression results depend on how closely they end up producing close estimations of the original data. For this reason, we can have excellent fits in the sense of the AIC criterion that are not very useful for fitting actual data because these original data have been damped excessively. As an example, consider Figure \ref{fig:tres_olivos}. Subfigures \ref{fig::tres_olivos_time_series_01} and \ref{fig::tres_olivos_time_series_09} show our regression results for two very disparate values of $\alpha$ ($0.1$ and $0.9$). In Subfigure \ref{fig::tres_olivos_time_series_01} the data filtered and splined is clearly wavier than in Subfigure \ref{fig::tres_olivos_time_series_09}, a consequence of a softer denoising effort. These wavier curves are very close to the original noisy data points and produce derivatives that are more difficult to fit using second-order polynomials, thus yielding results that do not adjust very good to the solid lines. In Subfigure \ref{fig::tres_olivos_time_series_09} the original time series has been modified much more significantly and the derivatives behave in a way that is more suitable for a second-order polynomial to be fitted. After integrating of the resulting system of ODEs, this results in dashed lines that effectively overlap the filtered and splined lines. However, our true measure of accuracy is given by the errors with respect to the original data, that is, the distance between the ``\texttimes" and the ``$\circ$" and, in that sense the scenario with $\alpha=0.1$ provides more accurate results for this station in this time window ($RMSE^{0.10}_{NO_2,Tres~Olivos}=0.3613$ vs $RMSE^{0.90}_{NO_2,Tres~Olivos}=0.5542$ and $RMSE^{0.10}_{O_3,Tres~Olivos}=0.2139$ vs $RMSE^{0.90}_{O_3, Tres~Olivos}=0.2588$). These differences are even more distinguishable in subfigures \ref{fig::tres_olivos_state_01} and \ref{fig::tres_olivos_state_09}, where the identified (simulated) trajectories are clearly better with respect to the original data points in the case with $\alpha=0.10$. 

\begin{figure}
     \centering
     \begin{subfigure}[h!]{0.49\textwidth}
         \centering
         \includegraphics[width=\textwidth]{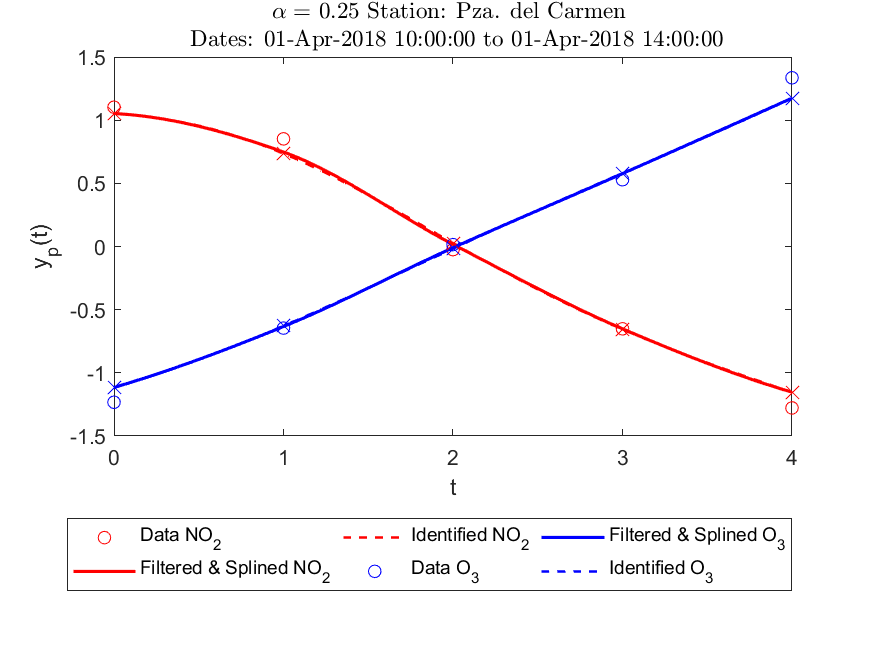}
         \caption{5-hour window}
         \label{fig:5h_time_series}
     \end{subfigure}
     \hfill
     \begin{subfigure}[h!]{0.49\textwidth}
         \centering
         \includegraphics[width=\textwidth]{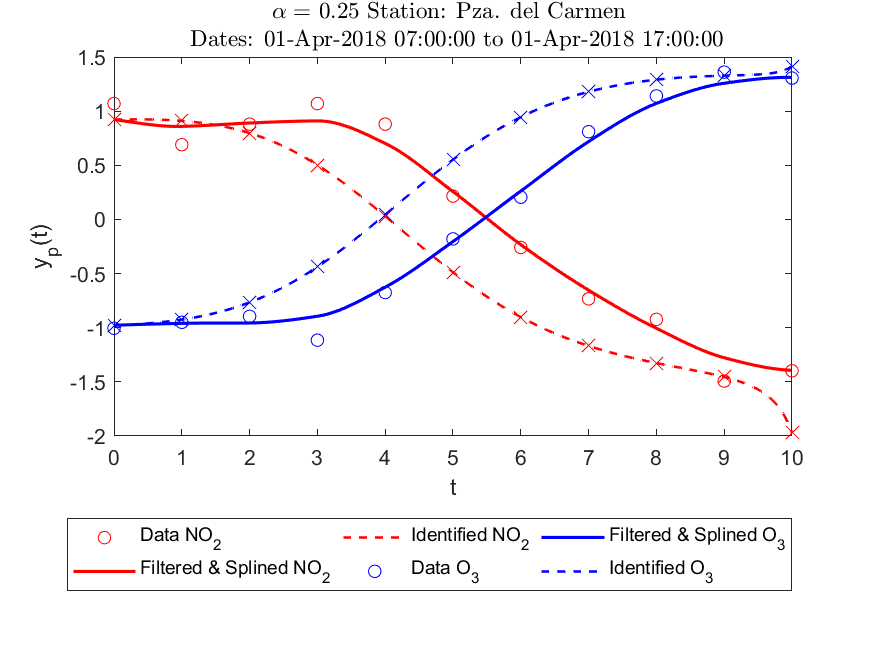}
         \caption{11-hour window}
         \label{fig:11h_time_series}
     \end{subfigure}
     \hfill
     \begin{subfigure}[h!]{1.0\textwidth}
     \centering
        \begin{tabular}{|c|c|c|c|c|c|c|c|}
            \hline
                Window & $\dot{y}_i$ & $\hat{\beta}_{{i}_0}$ & $\hat{\beta}_{{i}_1}$ & $\hat{\beta}_{{i}_2}$ & $\hat{\beta}_{{i}_3}$ & $\hat{\beta}_{{i}_4}$ & $\hat{\beta}_{{i}_5}$ \\\hline
                \multirow{2}{*}{\begin{turn}{0}5 hours\end{turn}} & $d[NO_2]/dt$ & -0.7561 & -1.2358 & -1.3949 & - & -1.0494 & -0.6539 \\
                & $d[O_3]/dt$ & -0.6142 & 0.6275 & 0.7181 & - & 0.2640 & 0.1814 \\\hline
                \multirow{2}{*}{\begin{turn}{0}11 hours\end{turn}} & $d[NO_2]/dt$ & -0.4279 & -0.7495 & -0.8854 & -6.3679 & -12.488 & -5.8383 \\
                & $d[O_3]/dt$ & 0.4317 & 0.9800 & 1.0719 & 2.7402 & 5.2943 & 2.2776 \\\hline
        \end{tabular}
        \caption{Fitted systems of differential equations under Akaike's Information Criterion for different time windows, $\alpha=0.25$. General form: $\dot{y}_{i}=\hat{\beta}_{{i}_0}+\hat{\beta}_{{i}_1}y_{NO_2}+\hat{\beta}_{{i}_2}y_{O_3}+\hat{\beta}_{{i}_3}y_{NO_2}^2+\hat{\beta}_{{i}_4}y_{NO_2}y_{O_3}+\hat{\beta}_{{i}_5}y_{O_3}^2, i=NO_2,O_3$. }
        \label{tab:fitted_systems_5_vs_11}
     \end{subfigure}
        \caption{Effect of the length of the time window on sparsity}
        \label{fig:length_and_sparsity}
\end{figure}

Another important result stems from the degree of sparsity found in the fitted regressions. The assumption that the time derivatives of the concentrations of chemical species in the troposphere can be modeled with second-order polynomials is, \textit {per se}, an assumption that specifically targets a type of functionals. With two chemical species like $NO_2$ and $O_3$, this left us with at most $6$ terms ($5$ variables and the intercept). From our results, it seemed that shorter time windows (and therefore most likely less noisy data sets) might produce sparser sets of ODEs under the AIC, ceteris paribus. This phenomenon is shown on Figure \ref{fig:length_and_sparsity}. However, in most occasions the solutions to Problem \eqref{eqn:best_subset_AIC} were full models or models without a large degree of sparsity. Therefore, we can say that while our intention was to use SINDy to further reduce the number of terms in these regressions, our results show that real-world data may not behave as well as generated data and hindered our ability to obtain sparser representations. In regards to the AIC, this means that the penalty imposed by the term $2\|\boldsymbol{\beta_i}\|_0$ is not sufficiently important to discard full models, even though in many instances we obtained very parsimonious models that performed almost as well as more complex regressions. In view of this, we also tried other criteria for selecting the best models such as the well-known $R^2_{adj}=1-(1-R^2)\frac{m}{(m-\|\boldsymbol{\beta_i}\|_0-1)}$ or the \textit{Bayesian Information Criterion} $BIC=m\log{(\frac{\|\mathbf{\dot{\tilde{y}}_i}-\mathbf{\tilde{F}}\boldsymbol{\beta_i}\|_2^2}{m})}+\log{(m)}(\|\boldsymbol{\beta_i}\|_0+1)$. We did not appreciate significant changes in the way these different methods ranked all the regressions. 

\section{Properties of Reconstructed ODEs}\label{sec:properties_ODEs}

In this section, we study some basic mathematical properties of the ODEs obtained from the data, which offer some further analytical insight about the dynamics of $[NO_2]$ and $[O_3]$. The purpose of this analysis is to highlight some tools that allow us to better interpret the resulting ODE models we obtained in the previous section.  
The general form of the  equations we are fitting is 
\begin{eqnarray}
\frac{d[NO_2]}{dt}=\hat{\beta}_{1_0}+\hat{\beta}_{1_1}[NO_2]+\hat{\beta}_{1_2}[O_3]+\hat{\beta}_{1_3}[NO_2]^2+
\hat{\beta}_{1_4}[O_3][NO_2]+\hat{\beta}_{1_5}[O_3]^2 \label{ode_NO2}\\
\frac{d[O_3]}{dt}=\hat{\beta}_{2_0}+\hat{\beta}_{2_1}[NO_2]+\hat{\beta}_{2_2}[O_3]+\hat{\beta}_{2_3}[NO_2]^2+
\hat{\beta}_{2_4}[O_3][NO_2]+\hat{\beta}_{2_5}[O_3]^2,\label{ode_O3}
\end{eqnarray}
where the coefficients $\{\hat{\beta}_{i_j}\}$ are obtained by solving the optimization Problem \eqref{eqn:best_subset_AIC}.


Such planar quadratic system
can exhibit diverse behaviors that are well-studied in the mathematical literature. For example, $[NO_2]$ can blow up in finite time if  $\hat{\beta}_{1_0}>0,\,\hat{\beta}_{1_3}>0$ and all other coefficients are zero. Phase plane analysis, which characterizes topological properties of the two-dimensional trajectories of the system, reveals that this system, in general, can describe more than $700$ different classes of phase portraits \citep{reyn2007phase}.


From the coefficients obtained in the time-window considered, $\hat{\beta}_{1_5}^2+\hat{\beta}_{2_5}^2$ and $\hat{\beta}_{1_3}^2+\hat{\beta}_{2_3}^2$ are both nonzero for all 14 stations. Furthermore, 
Theorems 2.1 and 2.2 of \cite{reyn2007phase} assert that the sum of the multiplicities (called \textit{finite multiplicity} $m_f$ in \cite{reyn2007phase}) of  the critical points of \eqref{ode_NO2}-\eqref{ode_O3} is 4 for all 14 stations. 
Below we summarize some results about real critical points\footnote{A real (respectively, complex) critical point is a critical point whose coordinates are all real (respectively, complex) numbers} for all 14 stations.

The system \eqref{ode_NO2}-\eqref{ode_O3} can be written in the compact form
\begin{equation}
    \dot{y}_1=P(y_1,y_2),\quad \dot{y}_2=Q(y_1,y_2),
\end{equation}
where $(y_1,y_2)=([NO_2], [O_3])$ and $P$ and $Q$ are quadratic polynomials. If  $P(y_1^{\ast}, y_2^{\ast} )=Q(y_1^{\ast}, y_2^{\ast} )=0$, then $(y_1^{\ast}, y_2^{\ast} )$ is called a critical point.  
The local stability of a  critical point $(y_1^{\ast}, y_2^{\ast} )$ is characterized by the eigenvalues   of $J(y_1^{\ast}, y_2^{\ast} )$, where 
\[
J(y_1,y_2)=
\begin{pmatrix}
\hat{\beta}_{1_0}+2\hat{\beta}_{1_3}y_1+\hat{\beta}_{1_4}y_2 & \hat{\beta}_{1_2}+2\hat{\beta}_{1_5}y_2+\hat{\beta}_{1_4}y_1\\
\hat{\beta}_{2_0}+2\hat{\beta}_{2_3}y_1+\hat{\beta}_{2_4}y_2 & \hat{\beta}_{2_2}+2\hat{\beta}_{2_5}y_2+\hat{\beta}_{2_4}y_1
\end{pmatrix}
\]
is the Jacobian matrix \citep[Section 2.3.1]{reyn2007phase} \citep[Appendix A]{murray2007mathematical}. 
For example, 
suppose the eigenvalues $\{\lambda_1,\lambda_2\}$ are real and distinct (without loss of generality $\lambda_1>\lambda_2$). Then the critical point is a stable node if $0>\lambda_1>\lambda_2$, an unstable node if $\lambda_1>\lambda_2>0$, and a saddle point if $\lambda>0>\lambda_2$. 

As mentioned in Section \ref{subsec:data}, the system \eqref{ode_NO2}-\eqref{ode_O3} was standardized by subtracting the average concentration  and dividing over the standard deviation. This normalization, being a linear transformation 
$$(y_1, y_2)\mapsto (w_1,w_2)=\left(\frac{y_1-\mu_1}{\sigma_1},\frac{y_2-\mu_2}{\sigma_2}\right),$$
where $\mu$'s are the means and $\sigma$'s are the standard deviations, will not change the stability of the critical point. Furthermore, $(w_1^{\ast}, w_2^{\ast} )$ is the critical point of the normalized system \eqref{ode_NO2}-\eqref{ode_O3} if and only if
$(\mu_1+w_1^{\ast}\sigma_1 ,\,\mu_2+w_2^{\ast}\sigma_2)$ is the critical point of the original (non-standardized) system.

We can apply these analysis to the ODEs obtained for the 14 stations. 
In summary, 9 stations have 4 real critical points, 4 stations have a pair of complex critical points and two real critical points, and  exactly one station has 4 complex critical points. 
Hence there are 44 real critical points and 12 complex critical points. 

Among the 44 real critical points, 36 have positive coordinates and deserve special attention since they have physical interpretation as chemical concentrations. One critical point has negative coordinates and the remaining
7 have one positive coordinate and one negative coordinate (Arturo Soria Station 
is an example). 
Stability analysis are performed for the 36 real critical points that have positive coordinates, 
which reveals that more than half (20 out of 36) are saddle points
(see Table \ref{table:stability} for a summary).
\begin{table}[h!]
\centering
\begin{tabular}{ |c|c|c| } 
 \hline
 \textbf{Type of critical point} & \textbf{Eigenvalues} $\{\lambda_1,\lambda_2\}$ & \textbf{Count} \\
 \hline\
Stable node & $0>\lambda_1>\lambda_2$ & 5 \\
Unstable node & $\lambda_1>\lambda_2>0$ & 4 \\ 
Saddle point& $\lambda_1>0>\lambda_2$ & 20 \\ 
Stable spiral & $\{a+bi ,a-bi\}$ where $a<0$ & 5 \\ 
Unstable spiral & $\{a+bi ,a-bi\}$ where $a>0$  & 2 \\
 \hline
\end{tabular}
\caption{Classification of the 36 real critical points with positive coordinates in terms of local stability.}
\label{table:stability}
\end{table}

\begin{figure}
     \centering
     \begin{subfigure}[h!]{0.49\textwidth}
         \centering
         \includegraphics[width=\textwidth]{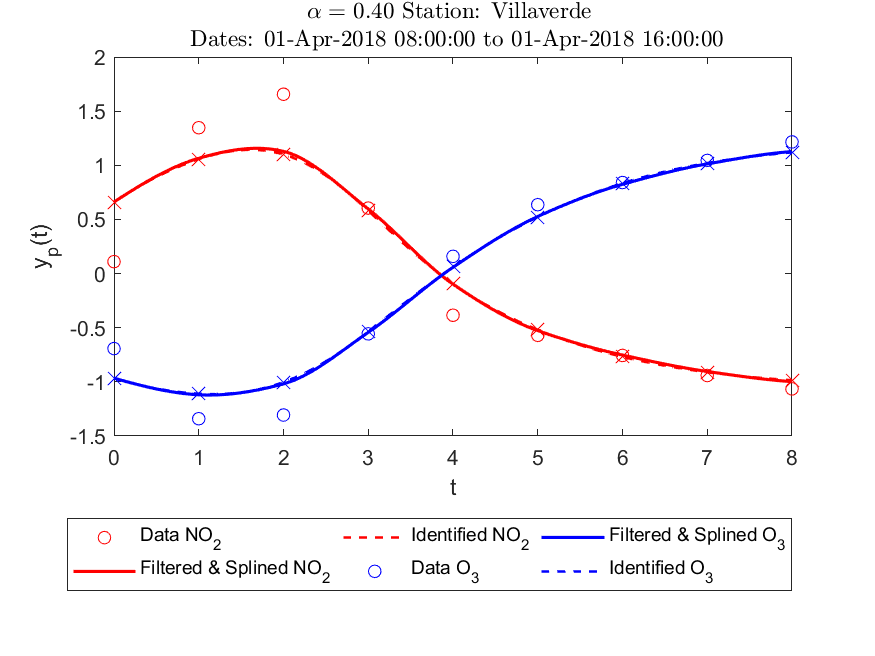}
         \caption{Time series}
         \label{fig::station17_a}
     \end{subfigure}
     \hfill
     \begin{subfigure}[h!]{0.49\textwidth}
         \centering
         \includegraphics[width=\textwidth]{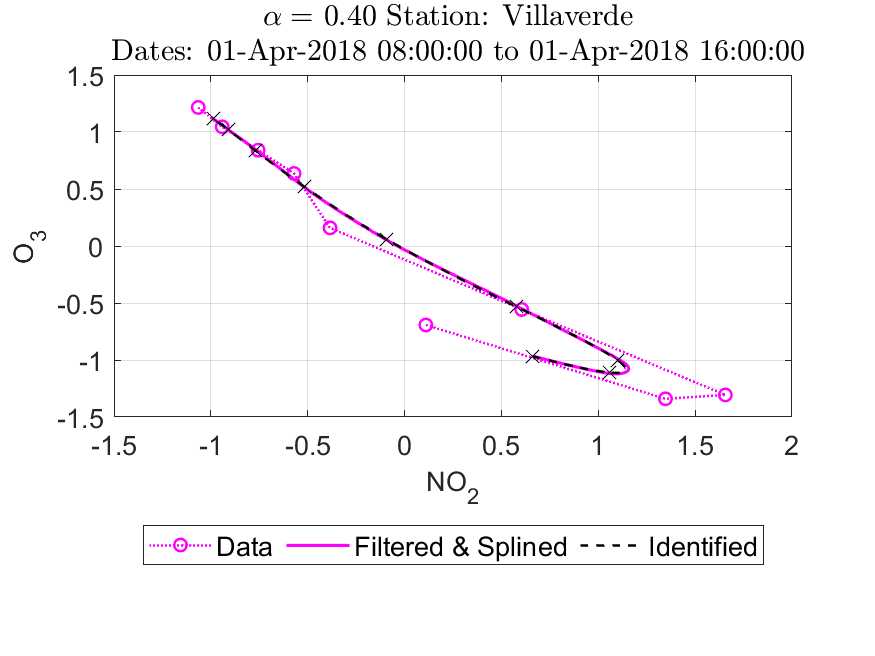}
         \caption{Two-dimensional trajectory}
         \label{fig::station17_b}
     \end{subfigure}
        \caption{(Left panel) Time series plots  for $NO_2$ and $O_3$ in Villaverde Station. 
        (Right Panel) The two-dimensional trajectory on the right starts from bottom-right and move towards top-left, getting closer to the stable node with standardized coordinates $(-1.0534, 1.2109)$.}
        \label{station28079017}
\end{figure}

We now give some examples to describe the behavior of the trajectory of the solution curves of the ODEs near the critical points in the $(y_1,y_2)$-plane. 
For example, there are  two  critical points for Villaverde Station. 
The critical point 
$(3.1894,91.8657)$, with standardized coordinates $(-1.0534,\, 1.2109)$,
is a  stable node  since  the  eigenvalues  $\{-4.4819, -0.7735\}$ of its  Jacobian  matrices are distinct and negative. A solution starting nearby this critical point will move towards and converge to the same critical point as time increases in the phase space (see Figure \ref{fig::station17_b}).
The other critical point $(81.4092,0.0714)$  in this station is a saddle point,
since  the  eigenvalues  $\{2.9546, -0.8807\}$ of its  Jacobian matrices have opposite signs. 
We do not have data near this critical point to validate the model around this point. In \ref{sec:appendix_critical}, we include figures for three other stations for further illustrations.
The critical point 
$(9.5075,\,82.7763)$ of Station Casa de Campo, with standardized coordinates 
$(-1.005
,\, 0.8465)$,
is a  stable spiral (see Figure \ref{fig:station28079024}).
The critical point 
$(52.9736
,\,13.1115)$ of Station Arturo Soria, with standardized coordinates 
$(1.5072,\, -1.2997)$,
is a  saddle point (see Figure \ref{fig:station28079016}). 
Station Farolillo has 4 critical points 
that are stable spiral, saddle point, unstable spiral and saddle point, respectively (see Figure \ref{fig:station28079018}).


%




Our method, when applied to longer time-windows, would give a planar quadratic system with time-varying coefficients $\{\hat{\beta}_{i_j}(t)\}$. Due to their simple form and rich structure, such closed form equations are promising tools for various purposes including the study of long-time stochastic behavior \citep{budhiraja2017uniform, nguyen2020stochastic} and the enhancement to spatial-temporal models such as  network models \citep{huang2010bayesian} and partial differential equations \citep{ogura1962scale, wilhelmson1972pressure, lanser1999analysis}.



\section{Reconstructing Missing Data}\label{sec:reconstruction}
\begin{figure}
\centering
  \begin{subfigure}[h!]{0.49\textwidth}
  \includegraphics[width=\textwidth]{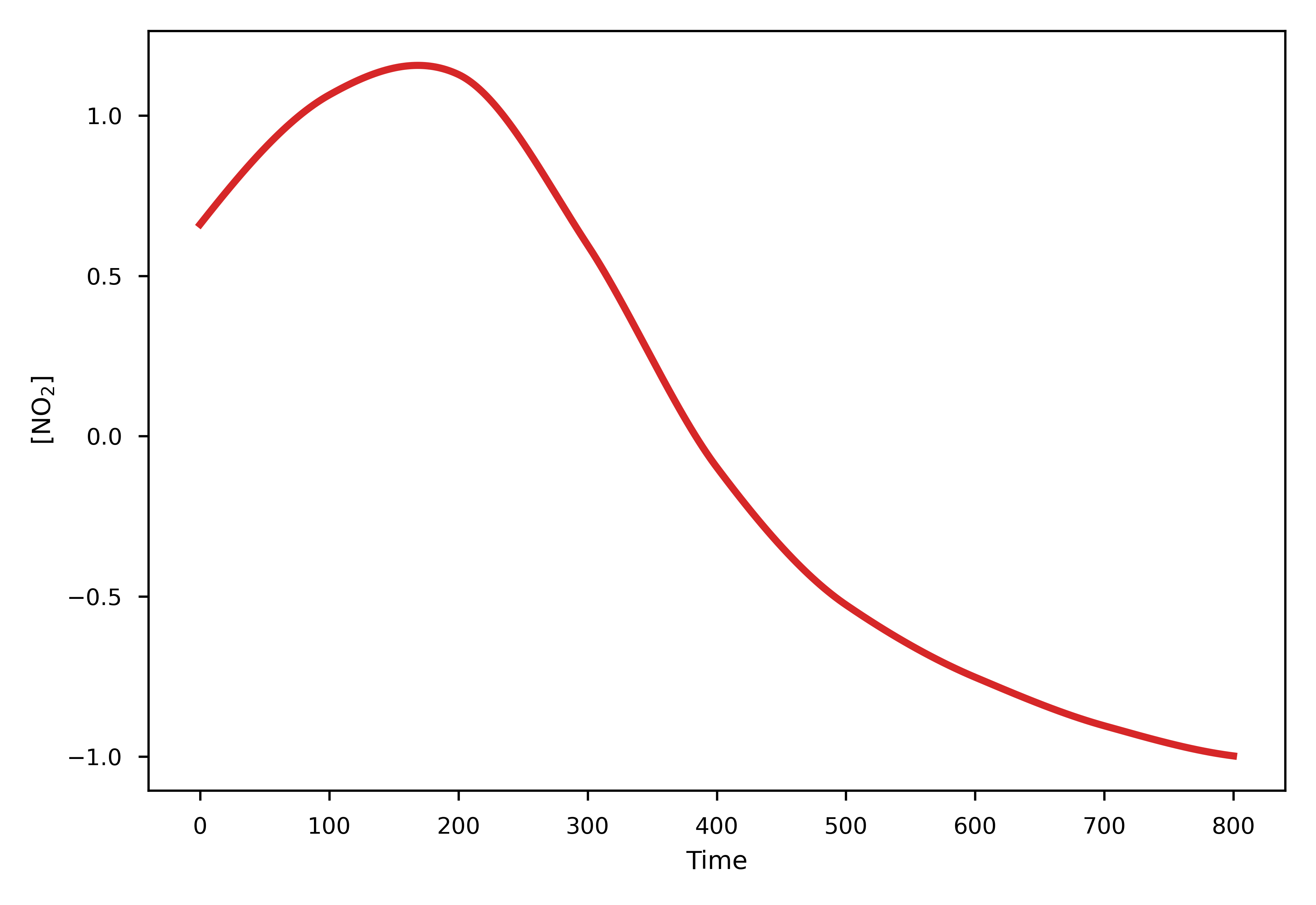}
  \caption{Normalized $[NO_2]$ Data for Villaverde Station. 
  }
  \label{fig:data}
  \end{subfigure}
  \hfill
  \begin{subfigure}[h!]{0.49\textwidth}
  \includegraphics[width=\textwidth]{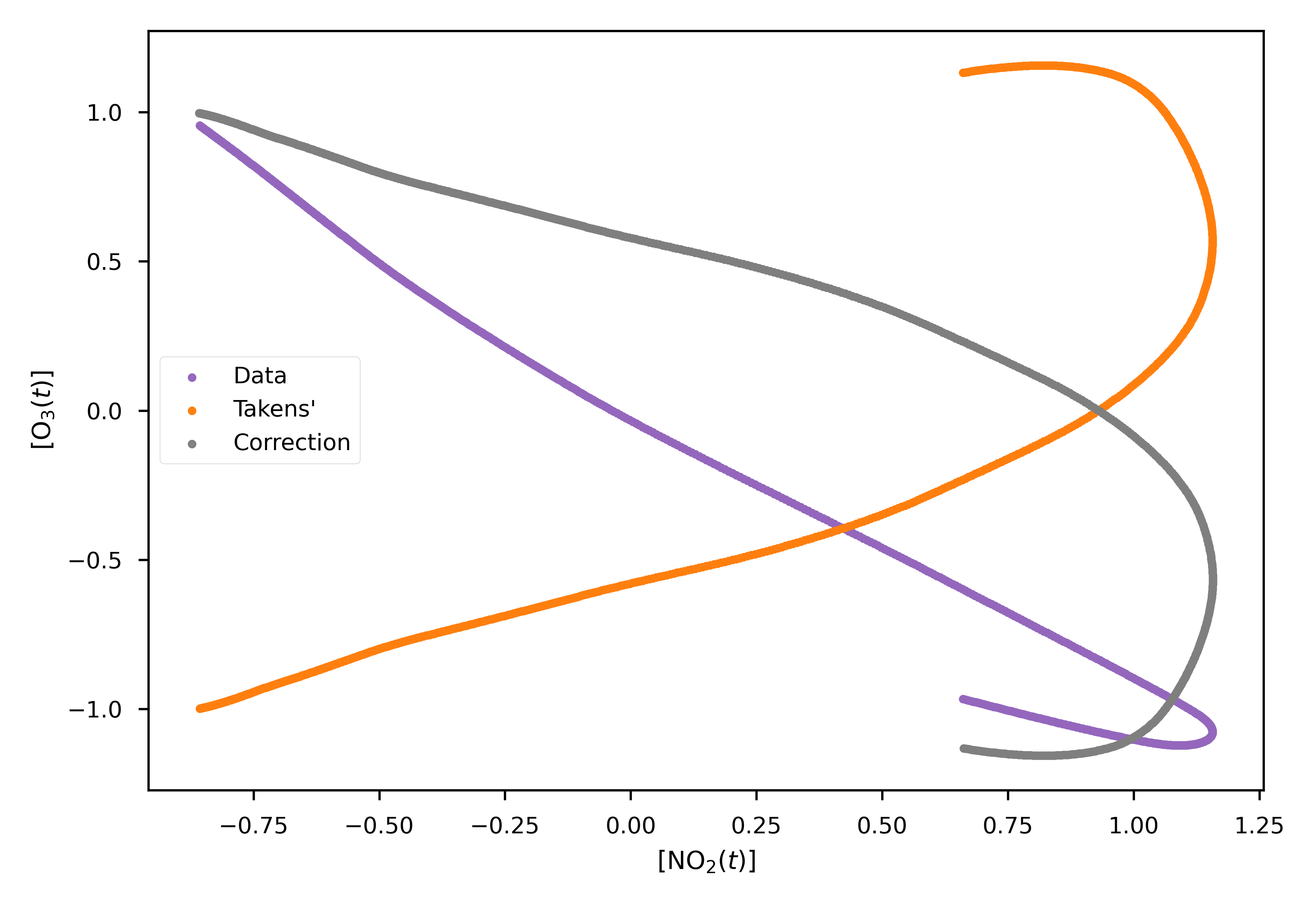}
  \caption{Reconstructed Trajectories.}
  \label{fig:take_reconstruction}
  \end{subfigure}
  \caption{Our goal is to recover as much qualitative information as possible about our
  underlying system using a single data coordinate. (\subref{fig:data}) Normalized $[NO_2]$ data for April $01, 2018$ from $8AM$ to $4PM$ at the Villaverde Station. (\subref{fig:take_reconstruction}) Comparison between the original trajectory (in purple), Takens' Reconstruction (in orange) with $\tau=136$, and Takens' output after optimizing over $O(2)$ (in grey).}  \label{fig:reconstruction}
\end{figure}

From our dataset, we were only able to fit 14 equations from 24 stations available. This is due to the fact that not all stations captured data for all the pollutants. As previously mentioned in Section \ref{subsec:data}, only 14 out of the 24 stations contained data for both $NO_2$ and $O_3$, but all 24 stations capture measurements for $NO_2$. Given our ability to fit the stations outlined in Section \ref{sec:SINDy}, we hypothesize that we can use Takens' delay embedding theorem \citep{takens1981detecting} to recover the $O_3$ measurements from our data. In 1981, Floris Takens showed that global features of a trajectory in a dynamical system can be recovered using a single coordinate from the original data. In practice the following map,
\begin{equation}\label{lag_map}
\begin{split}
\hat{\mathbf{y}}(t)=\Phi_{\tau, d}(\mathbf{y}(t))=(y_1(t),\ &y_1(t+\tau),\ y_1(t+2\tau),\ldots,\ y_1(t+(d-1)\tau))
\end{split}
\end{equation}
yields an embedding that recovers the global properties of the original trajectory, for some lag $\tau$ and embedding dimension $d$. We represent $y_1$ as the first coordinate of the original data vector ${\mathbf{y}}$. If we assume that the data collected can be linked by an ODE using SINDy, as demonstrated in Section \ref{sec:SINDy}, we should be able to recover partial information from our missing $O_3$ measurements\footnotemark.
\footnotetext{See \citep{sauer1991embedology} for a detailed account of the necessary and sufficient conditions that need to be meet before the results from Takens' delay embedding theorem can be applied. These conditions hold for the equations we obtained when studying the Villaverde station in Figure \ref{fig:reconstruction}.}
In practice, we need to estimate both $\tau$ and $d$, but in our case we can restrict ourselves to $d=2$ since we are considering $NO_2$ and $O_3$. To estimate $\tau$, we use the method of minimizing the average mutual information between the data and the first lag coordinate (or second coordinate of  equation \eqref{lag_map}) \citep{fraser1986independent}.

Define the average mutual information of a time series $\mathbf{\tilde{y}}=(y(t_1), y(t_2),\dots, y(t_m))$ with lag
$\tau$ by 
\begin{equation*}
AMI(\tau)=\sum_{i,j}q_{ij}(\tau)\log\left(\frac{q_{ij}(\tau)}{q_iq_j}\right),
\end{equation*}
where $q_i$  is the probability that $y(t_l)$ is in bin $i$ of the histogram
constructed using samples from $\mathbf{\tilde{y}}$, and $q_{ij}(\tau)$ is the probability
that $y(t_l)$ is in bin $i$ and $y(t_l + \tau)$ is in bin $j$ \citep{wallot2018calculation}.


 As mentioned previously, the embedding $\hat{\mathbf{y}}(t)$ will only recover some qualitative features of the trajectory, but for our problem we can exploit the fact that we only have two dimensions to recover some information about the geometry of the trajectory (e.g. the asymptotic behavior). Consider the following optimization problem 
\begin{equation}
\label{opt_rot}
A = \displaystyle{\argmin_{R\in \mathrm{O}(2)} \sum_{t} |P_{y_1}R\hat{\mathbf{y}}(t) - y_1(t)|^2 },
\end{equation}
where $O(2)$ is the orthogonal group of all $2\times2$ matrices and $P_{y_1}$ is the projection onto the first coordinate (our data coordinate). Exploring the local minima of this loss landscape yields matrices $A$ (excluding the trivial solution $A=I$) that can be used to transform the output obtained from the reconstruction by applying $A\hat{\mathbf{y}}(t)$. Figure \ref{fig:reconstruction} shows the results we obtained when performing Takens' reconstruction and its correction for the Villaverde Station using $NO_2$ measurements to reconstruct $O_3$ samples (see \ref{sec:appendix_reconstruction} for more details). It is worth mentioning that the ODE that SINDy recovered from the data shown in Figure \ref{fig:reconstruction} is near an attractor, in this case a stable point, and this is a sufficient condition for the reconstruction procedure to apply \citep{sauer1991embedology}.

This technique can help engineers qualitatively recover pollutant measurements that were originally not captured without the need to upgrade equipment. Further exploration needs to be done in order to properly reconstruct the geometry of the trajectories.


\section{Conclusion}

We have validated and outlined a series of data-driven tools to deal with real-world atmospheric time series data and showed how SINDy provides a framework for extracting ODE models for real data collected across multiple stations distributed throughout the city of Madrid. We can break down our conclusions at three different levels:

\begin{enumerate}
    \item Descriptive analysis: We unveiled that using LASSO for extracting parsimonious ODEs in the context of SINDy can present numerical issues when solving these equations. We discussed how best subset regression, along with the Akaike information criterion, offered us a more stable way to consistently fit multiple stations and how they can be combined with an optimization framework that selects the optimal level of noise dampening as to produce the best possible fit with respect to our original data. We find that, as we aim at identifying the dynamics of our system for longer periods of time, this dampening has to be more intense and a good combination of sparsity and goodness of fit is more difficult to attain.
    \item Stability analysis: We performed stability analysis of the reconstructed ODEs in order to highlight global features of the space of possible trajectories and provide us with insight of how the concentrations of the chemical species under consideration may change over time beyond the period of study. This analysis is based on the idealized assumption that all environmental conditions, except the concentrations of $NO_2$ and $O_3$, are constant over time (i.e., the coefficients $\hat{\beta}_{i_j}$ are constants). We found that more than half among the physically relevant critical points are saddle points, suggesting that the system is unstable even under idealized environmental assumptions. However, there are few stable critical points in the model, pointing to a discrepancy with observed data in a longer ($>$ 24 hours) time-scale. This discrepancy suggests that  future refinements of the model can involve time-inhomogeneous coefficients  that capture environmental fluctuations.
     
   \item Reconstruction of trajectories: We discuss a reconstruction technique using Takens' embedding theorem that allows for the recovery of missing pollutant concentration data using other correlated concentration measurements. We show how we can reconstruct qualitatively $O_3$ measurements by only using $NO_3$ measurements. We suggest the use of matrix transformations as a way to correct for unfeasible solutions obtain from Takens' Reconstruction. This technique can be useful to use at stations that are not equipped to measure all atmospheric pollutants.
\end{enumerate}

In short, our methodology will provide researchers with the ability to construct interpretable, data-driven surrogate models from noisy chemical data sets. More importantly, it provides a more complete picture of the behavior of real world atmospheric chemical species ($NO_2$ and $O_3$ in our case, although our methodology can be extended to any others). We hope that the results obtained from the wide adoption of these tools allow pertinent authorities and policy makers make more informed decisions when designing future environmental policies. 

\section*{Acknowledgment}
The authors want to thank Michael S. Hughes and Paul Bruillard for helpful discussion on Takens' theorem and the theory of dynamical systems. W.T. Fan gratefully acknowledge the support of NSF grant DMS-1804492 and ONR grant TCRI N00014-19-S-B001. Support for C. Ortiz Marrero was provided by the Laboratory Directed Research and Development Program at Pacific Northwest National Laboratory, a multi-program national laboratory operated by Battelle for the U.S. Department of Energy, Release No. PNNL-SA-157007.

\Urlmuskip=0mu plus 1mu\relax
\bibliographystyle{apa}
\bibliography{./bibliography/bibliography}



\newpage

\appendix
\section{Methods considered}\label{sec:methods_considered}

\begin{enumerate}
\item Best subset regression \citep[Chapter~19]{hill2006statistics}: The very notion of sparse regression suggests the selection of a subset of $1\le c\le n$ terms. This subproblem, called the \textit{best subset} problem, can be cast generally as:

\begin{mini}|l|[1]
  {\boldsymbol{\beta_i}}{f(\boldsymbol{\beta_i})}{}{}
  {\label{eqn:best_subset}}
\addConstraint{\|\boldsymbol{\beta_i}\|_0}{\le c,}{}
\end{mini}
\noindent where the $\ell_0$-norm $\|\boldsymbol{\beta_i}\|_0=\sum_{j=1}^{n} 1\{\beta_{i{_j}}\neq 0\}$ is an indicator function that denotes the number of non-zero elements of the vector $\boldsymbol{\beta_i}$ (except for the intercept). Therefore, Problem \eqref{eqn:best_subset} aims at finding a sparse representation of $\dot{y}_i(t)$ that has at most $c$ terms and that minimizes $f(\boldsymbol{\beta_i})$ (very often the sum of errors squared, i.e., $f(\boldsymbol{\beta_i})=\frac{1}{2}\|\mathbf{\dot{\tilde{y}}_i}-\mathbf{\tilde{F}}\boldsymbol{\beta_i}\|_2^2$, also known as \textit{least-squares regression}). However, its only constraint is combinatorial in nature and makes this optimization problem NP-hard \citep{natarajan1995sparse}. Thus, despite of very promising and recent efforts with mixed-integer optimization reformulations \citep{bertsimas2016best}, researches and practitioners alike usually resort to different alternatives to attain sparse regressions.
\item LASSO regression \citep{tibshirani1996regression}: One such option lies on a convex quadratic alternative to Problem \eqref{eqn:best_subset} known as \textit{LASSO} (\textbf{L}east \textbf{A}bsolute \textbf{S}hrinkage and \textbf{S}election \textbf{O}perator) regression:
\begin{mini*}|l|[1]
  {\boldsymbol{\beta_i}}{\frac{1}{2}\|\mathbf{\dot{\tilde{y}}_i}-\mathbf{\tilde{F}}\boldsymbol{\beta_i}\|_2^2}{}{}
  \addConstraint{\|\boldsymbol{\beta_i}\|_1}{\le\phi,}
\end{mini*}
\noindent where $\|\boldsymbol{\beta_i}\|_1=\sum_{j=1}^{n}|\beta_{i_j}|$ is the $\ell_1$-norm of the vector $\boldsymbol{\beta_i}$ (except the intercept). If we denote by $\boldsymbol{\hat{\beta}_i}^*$ the values of the regression coefficients of the full (i.e., the unconstrained) regression, then any value of $\phi$ such that $\|\boldsymbol{\hat{\beta}_i}^*\|_1>\phi$ will produce a shrinkage. Geometrical considerations in this model make this shrinkage such that some coefficients will be identical to zero as we decrease the upper bound on the $\ell_1$-norm (see \citep{tibshirani1996regression} for more details). This optimization model is frequently expressed as an equivalent unconstrained problem with a regularization parameter $\lambda$:
\begin{mini}|l|[1]
  {\boldsymbol{\beta_i}}{\frac{1}{2}\|\mathbf{\dot{\tilde{y}}_i}-\mathbf{\tilde{F}}\boldsymbol{\beta_i}\|_2^2+\lambda\|\boldsymbol{\beta_i}\|_1}{}{LR(i):}
  {\label{eqn:lasso_2}}
\end{mini}
Since the problems $LR(i), i=1,2,\dots,p$ are quadratic and convex, they can be efficiently solved by some well-known optimization methods for finding the optimal solutions of a convex quadratic function over a polyhedron (see \citep{nocedal2006numerical} for a reference of some usual optimization methods for this kind of problems). It is worth mentioning that many researchers have historically seen Problem \eqref{eqn:lasso_2} as a heuristic to solve Problem \eqref{eqn:best_subset}, which is widely regarded as the formulation that yields the most desired sparse solution with a subset of $c$ variables. However, as noted in \citep{hastie2017extended}, in noisy settings both problems offer different bias-variance tradeoffs and, for this reason, the superiority of best subset regression over LASSO regression is not clear-cut. 
\end{enumerate}

\newpage
\section{Algorithm for SINDy with AIC for one station}\label{sec:appendix_algorithm}

\begin{algorithm*}[h!]
\SetAlgoLined
\KwResult{Find the best regression in the sense of AIC that did not present singularities.}
 Let $p$ be the number of chemical species; $n$ be the number of variables in the full regression model; $\epsilon$ be a threshold for the maximum value allowed for $\dot{y}_i(t)$\; $l_i=1$ be a counter denoting which ranked model for chemical species $i$ should be selected\; $l=0$ be a counter for the total number of models discarded\;

 Find the $\mathcal{F}=\bigcup_{j=1}^{2^n}\mathcal{F}_j$ possible subsets of variables with size less or equal to $n$\;
 \For{$i\gets1$ \KwTo $p$}{
    \For{$j\gets1$ \KwTo $2^n$}{
         Solve the system (\ref{eqn:regression}) in the sense of least-squares with the subset of variables $\mathcal{F}_j$\;
    }
     Rank the $2^n$ different models for the $i^{th}$ chemical species according to the AIC\;
}

 \While {$l\ge0$}{
     \For{$i\gets1$ \KwTo $p$}{
     Select the $l_{i}^{th}$ best-ranked model for chemical species $i$\;
     }
    With the selected models, solve numerically the system of ODEs defined in (\ref{ode_NO2}-\ref{ode_O3})\;
    \eIf{$\dot{y}_i(t)>\epsilon$}{
        $l_i=l_i+1$\;
        $l=l+1$\;
        }{
        $l=-1$\;
    }
 }
\caption{SINDy with AIC for one station}
\end{algorithm*}

\newpage
\section{Reconstruction Algorithm}\label{sec:appendix_reconstruction}
\setcounter{figure}{0}

\begin{figure}[htbp] 
\centering
 \includegraphics[width=0.8\textwidth,height=0.8\textheight,keepaspectratio]{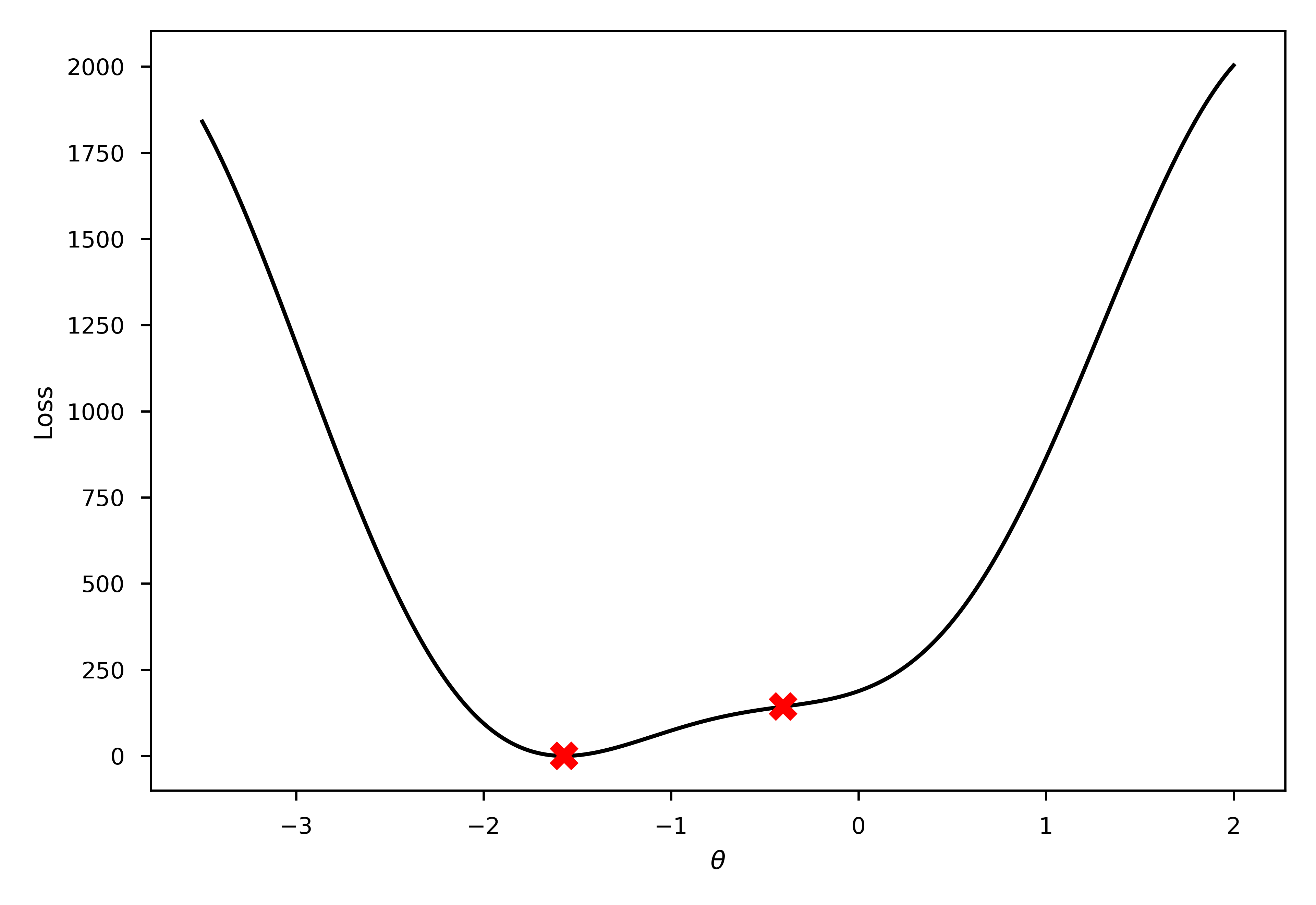}
 \caption{This figure represents a portion of the loss landscape when minimizing equation \ref{opt_rot} using the data outlined in Figure \ref{fig:reconstruction}. In order to visualize the loss landscape, we use the fact that $2\times 2$ rotations and reflections can be parametrized by one angle. Consider the following family of parametrized matrices in $O(2)$,
 $\big(\begin{smallmatrix}
 0 & 1 \\
 1 & 0 
\end{smallmatrix}\big)\big(\begin{smallmatrix}
\cos(\theta) & \sin(\theta) \\
-\sin(\theta) & \cos(\theta) 
\end{smallmatrix}\big)$ for some angle $\theta$.
 The marked values correspond to local minima of equation \ref{opt_rot} over this family of matrices. The far left marked value corresponds to the angle we chose to perform the correction i.e. the grey curve show in Figure \ref{fig:reconstruction}. 
 \label{opt_loss}
 }

\end{figure}

Figure \ref{opt_loss} contains more details on the loss landscape we explored to obtain the reconstructed trajectory in Figure \ref{fig:take_reconstruction}. In higher dimensions, the optimization loss landscape becomes difficult to visualize and the problem becomes ill-posed as we increase the dimensionality and decrease the access to data. Nevertheless, there are packages such as \citep{townsend2016pymanopt} that allow users to solve this optimization problem over sets of $n\times n$ matrices and this approach could shed some light into potential reconstructions for higher dimensional problems.

\section{Further examples of phase portraits}\label{sec:appendix_critical}
\setcounter{figure}{0}

In addition to Figure \ref{station28079017}, we give three more examples to visualise the time series data at different stations together with the critical points of the reconstructed ODEs.

\begin{figure}
     \centering
     \begin{subfigure}[h!]{0.49\textwidth}
         \centering
         \includegraphics[width=\textwidth]{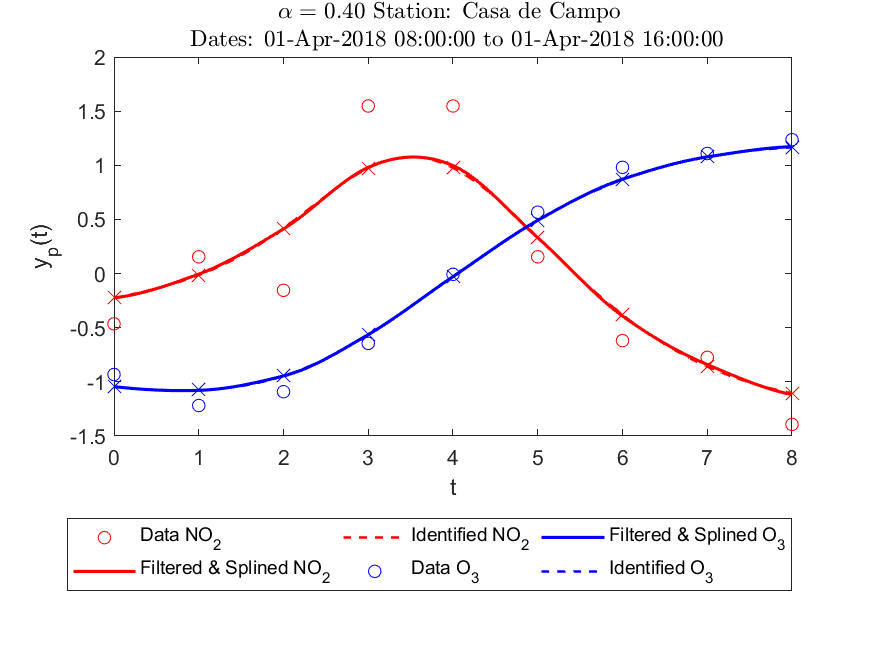}
         \label{fig::station24_a}
     \end{subfigure}
     \hfill
     \begin{subfigure}[h!]{0.49\textwidth}
         \centering
         \includegraphics[width=\textwidth]{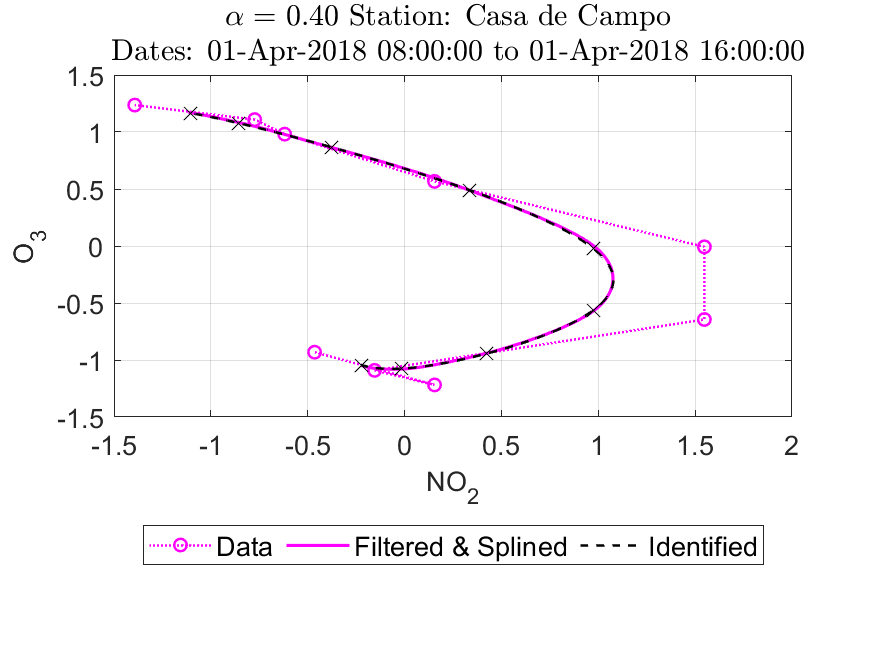}
         \label{fig::station24_b}
     \end{subfigure}
    \vspace{-4 em}
        \caption{(Left panel) Time series plots  for $NO_2$ and $O_3$ in Casa de Campo Station. 
        (Right Panel)  Two-dimensional trajectory plot. The standardized state $(-1.005
,\, 0.8465)$ 
is a stable spiral.}
        \label{fig:station28079024}

\vspace{1em}

     \centering
     \begin{subfigure}[h!]{0.49\textwidth}
         \centering
         \includegraphics[width=\textwidth]{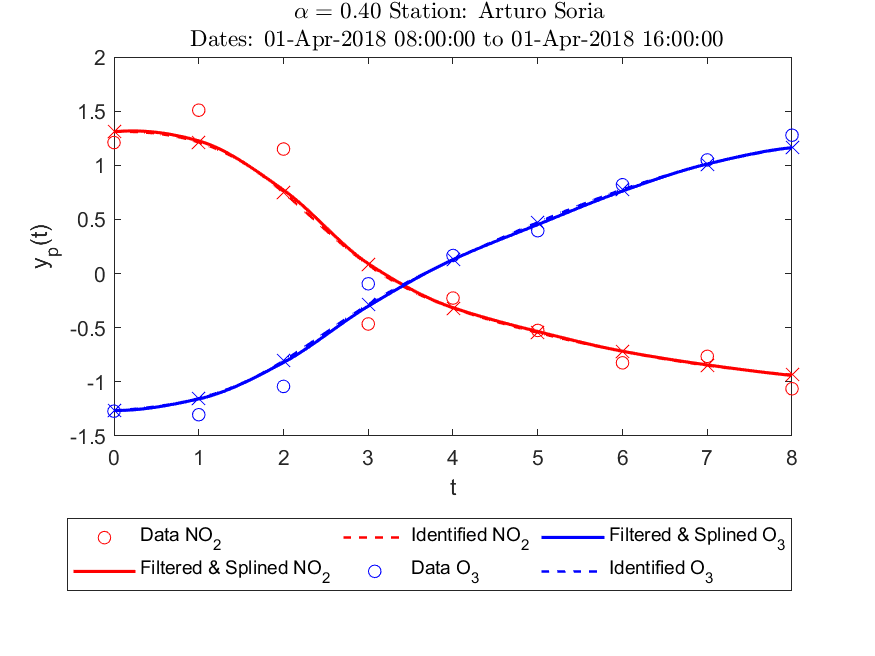}
     \end{subfigure}
     \hfill
     \begin{subfigure}[h!]{0.49\textwidth}
         \centering
         \includegraphics[width=\textwidth]{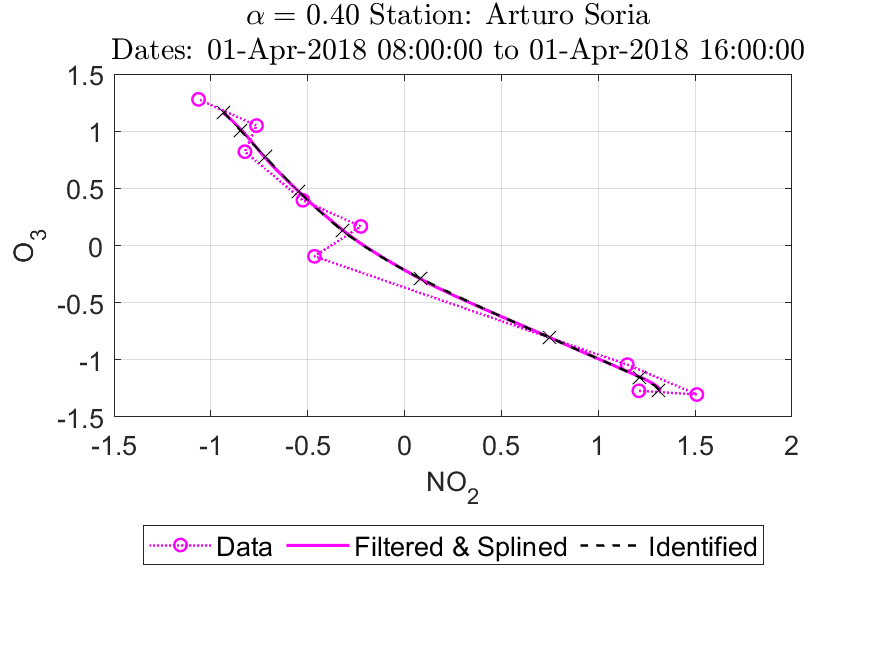}
     \end{subfigure}
    \vspace{-2.5 em}
        \caption{(Left panel) Time series plots  for $NO_2$ and $O_3$ in Arturo Soria Station. 
        (Right Panel)  Two-dimensional trajectory plot. The standardized state $(1.5072,\, -1.2997)$
is a  saddle point.}
        \label{fig:station28079016}

\vspace{1em}

     \centering
     \begin{subfigure}[h!]{0.49\textwidth}
         \centering
         \includegraphics[width=\textwidth]{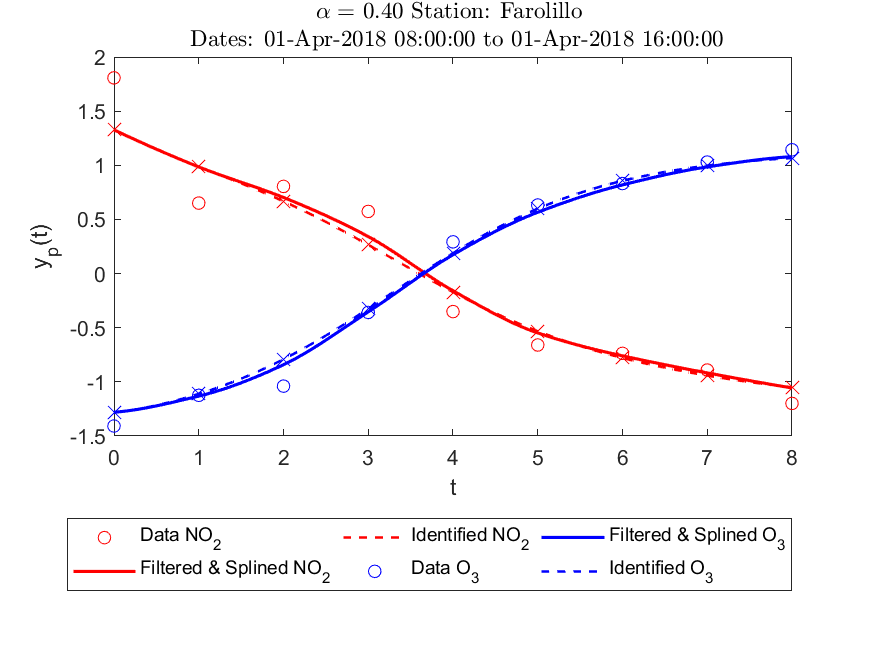}
     \end{subfigure}
     \hfill
     \begin{subfigure}[h!]{0.49\textwidth}
         \centering
         \includegraphics[width=\textwidth]{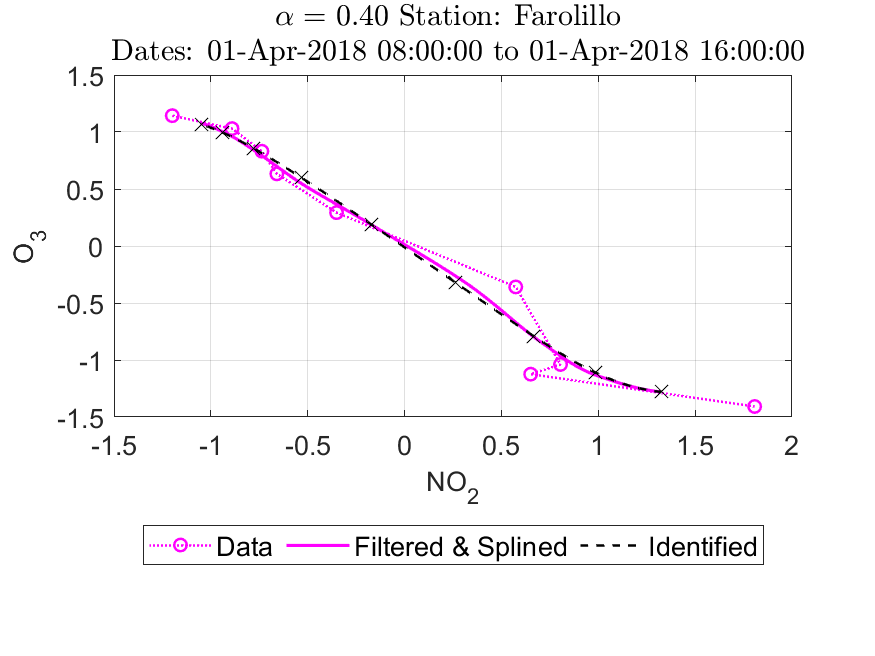}
     \end{subfigure}
    \vspace{-2.5 em}
        \caption{(Left panel) Time series plots  for $NO_2$ and $O_3$ in Farolillo Station. 
        (Right Panel) Two-dimensional trajectory plot. The 
     4 critical points with standardized coordinates $(-1.1518,1.0645)$, $(1.1669,-1.4159)$, $(0.6054,-0.2087)$ and $(-0.6740,0.9399)$ are stable spiral, saddle point, unstable spiral and saddle point respectively.}
        \label{fig:station28079018}
\end{figure}


\end{document}